\documentclass[twocolumn]{aastex63}
\usepackage{graphicx}
\usepackage{float}
\usepackage[T1]{fontenc}
\usepackage{aecompl}
\usepackage[]{amsmath}
\usepackage{color}
\usepackage{xspace}
\usepackage{soul}
\usepackage{booktabs} 
\usepackage{array} 
\usepackage{multirow} 
\usepackage{orcidlink}
\usepackage{tabularx}
\usepackage{lineno}

\graphicspath{{figures/}}

\def\lsim{ \lower .75ex \hbox{$\sim$} \llap{\raise .27ex \hbox{$<$}} }

\submitjournal{ApJ}

\shorttitle{DESI Y1 distance measurements}
\shortauthors{Li et al.}
\graphicspath{{./}{figures/}}

\begin{document}

\title{SpecDis: Value added distance catalog for 4 million stars from DESI Year-1 data}

\author{Songting Li}
\thanks{songtingli@sjtu.edu.cn}
\affiliation{Department of Astronomy, Shanghai Jiao Tong University, Shanghai 200240, China}
\affiliation{Shanghai Key Laboratory for Particle Physics and Cosmology, Shanghai 200240, China}
\affiliation{State Key Laboratory of Dark Matter Physics, School of Physics and Astronomy,Shanghai Jiao Tong University, Shanghai 200240, China}
\author{Wenting Wang}
\thanks{corresponding author: wenting.wang@sjtu.edu.cn}
\affiliation{Department of Astronomy, Shanghai Jiao Tong University, Shanghai 200240, China}
\affiliation{Shanghai Key Laboratory for Particle Physics and Cosmology, Shanghai 200240, China}
\affiliation{State Key Laboratory of Dark Matter Physics, School of Physics and Astronomy,Shanghai Jiao Tong University, Shanghai 200240, China}
\author{Sergey E. Koposov}
\thanks{Sergey.Koposov@ed.ac.uk}
\affiliation{Institute for Astronomy, University of Edinburgh, Royal Observatory, Blackford Hill, Edinburgh EH9 3HJ, UK}
\affiliation{Institute of Astronomy, University of Cambridge, Madingley Road, Cambridge CB3 0HA, UK}
\author{Ting S. Li}
\affiliation{Department of Astronomy and Astrophysics, University of Toronto, 50 St. George Street, Toronto ON, M5S 3H4, Canada}
\author{Youjia Wu}
\affiliation{Department of Astronomy, University of Michigan, Ann Arbor, MI 48109, USA}
\affiliation{Leinweber Center for Theoretical Physics, University of Michigan, 415 Church Street, Ann Arbor, MI 48105, USA}
\author{Monica Valluri}
\thanks{mvalluri@umich.edu}
\affiliation{Department of Astronomy, University of Michigan, Ann Arbor, MI 48109, USA}
\affiliation{University of Michigan, 500 S. State Street, Ann Arbor, MI 48109, USA}
\author{Joan Najita}
\affiliation{NSF NOIRLab, 950 N. Cherry Ave., Tucson, AZ 85719, USA}
\author{Carlos Allende Prieto}
\affiliation{Departamento de Astrof\'{\i}sica, Universidad de La Laguna (ULL), E-38206, La Laguna, Tenerife, Spain}
\affiliation{Instituto de Astrof\'{\i}sica de Canarias, C/ V\'{\i}a L\'{a}ctea, s/n, E-38205 La Laguna, Tenerife, Spain}
\author{Amanda Bystr{\"o}m}
\affiliation{Institute for Astronomy, University of Edinburgh, Royal Observatory, Blackford Hill, Edinburgh EH9 3HJ, UK}
\author{Christopher J. Manser}
\affiliation{Astrophysics Group, Department of Physics, Imperial College London, Prince Consort Rd, London, SW7 2AZ, UK}
\affiliation{Department of Physics, University of Warwick, Gibbet Hill Road, Coventry, CV4 7AL, UK}
\author{Jiaxin Han}
\affiliation{Department of Astronomy, Shanghai Jiao Tong University, Shanghai 200240, China}
\affiliation{Shanghai Key Laboratory for Particle Physics and Cosmology, Shanghai 200240, China}
\affiliation{State Key Laboratory of Dark Matter Physics, School of Physics and Astronomy,Shanghai Jiao Tong University, Shanghai 200240, China}
\author{Carles G. Palau}
\affiliation{Department of Astronomy, Shanghai Jiao Tong University, Shanghai 200240, China}
\affiliation{Shanghai Key Laboratory for Particle Physics and Cosmology, Shanghai 200240, China}
\affiliation{State Key Laboratory of Dark Matter Physics, School of Physics and Astronomy,Shanghai Jiao Tong University, Shanghai 200240, China}
\author{Hao Yang}
\affiliation{Tsung-Dao Lee Institute, Shanghai Jiao Tong University, Shanghai, 201210, China}
\affiliation{Department of Astronomy, Shanghai Jiao Tong University, Shanghai 200240, China}
\affiliation{Shanghai Key Laboratory for Particle Physics and Cosmology, Shanghai 200240, China}
\affiliation{State Key Laboratory of Dark Matter Physics, School of Physics and Astronomy,Shanghai Jiao Tong University, Shanghai 200240, China}
\author{Andrew P. Cooper}
\affiliation{Institute of Astronomy and Department of Physics, National Tsing Hua University, 101 Kuang-Fu Rd. Sec. 2, Hsinchu 30013, Taiwan}
\affiliation{Center for Informatics and Computation in Astronomy, NTHU, 101 Kuang-Fu Rd. Sec. 2, Hsinchu 30013, Taiwan}
\author{Namitha Kizhuprakkat}
\affiliation{Institute of Astronomy and Department of Physics, National Tsing Hua University, 101 Kuang-Fu Rd. Sec. 2, Hsinchu 30013, Taiwan}
\affiliation{Center for Informatics and Computation in Astronomy, NTHU, 101 Kuang-Fu Rd. Sec. 2, Hsinchu 30013, Taiwan}
\author{Alexander H.~Riley}
\affiliation{Institute for Computational Cosmology, Department of Physics, Durham University, South Road, Durham DH1 3LE, UK}
\author{Leandro Beraldo e Silva}
\affiliation{Department of Astronomy, University of Michigan, Ann Arbor, MI 48109, USA}
\affiliation{Steward Observatory, University of Arizona, 933 N, Cherry A ve, T ucson, AZ 85721, USA }
\author{Jessica Nicole Aguilar}
\affiliation{Lawrence Berkeley National Laboratory, 1 Cyclotron Road, Berkeley, CA 94720, USA}
\author{Steven Ahlen}
\affiliation{Physics Dept., Boston University, 590 Commonwealth Avenue, Boston, MA 02215, USA}
\author{David Bianchi}
\affiliation{Dipartimento di Fisica ``Aldo Pontremoli'', Universit\`a degli Studi di Milano, Via Celoria 16, I-20133 Milano, Italy}
\affiliation{INAF-Osservatorio Astronomico di Brera, Via Brera 28, 20122 Milano, Italy}
\author{David Brooks}
\affiliation{Department of Physics \& Astronomy, University College London, Gower Street, London, WC1E 6BT, UK}
\author{Todd Claybaugh}
\affiliation{Lawrence Berkeley National Laboratory, 1 Cyclotron Road, Berkeley, CA 94720, USA}
\author{Axel de la Macorra}
\affiliation{Instituto de F\'{\i}sica, Universidad Nacional Aut\'{o}noma de M\'{e}xico,  Circuito de la Investigaci\'{o}n Cient\'{\i}fica, Ciudad Universitaria, Cd. de M\'{e}xico  C.~P.~04510,  M\'{e}xico}
\author{John Della Costa}
\affiliation{NSF NOIRLab, 950 N. Cherry Ave., Tucson, AZ 85719, USA}
\affiliation{Department of Astronomy, San Diego State University, 5500 Campanile Drive, San Diego, CA 92182, USA}
\author{Arjun Dey}
\affiliation{NSF NOIRLab, 950 N. Cherry Ave., Tucson, AZ 85719, USA}
\author{Peter Doel}
\affiliation{Department of Physics \& Astronomy, University College London, Gower Street, London, WC1E 6BT, UK}
\author{Jaime E. Forero-Romero}
\affiliation{Departamento de F\'isica, Universidad de los Andes, Cra. 1 No. 18A-10, Edificio Ip, CP 111711, Bogot\'a, Colombia}
\affiliation{Observatorio Astron\'omico, Universidad de los Andes, Cra. 1 No. 18A-10, Edificio H, CP 111711 Bogot\'a, Colombia}
\author{Enrique Gazta\~naga}
\affiliation{Institut d'Estudis Espacials de Catalunya (IEEC), c/ Esteve Terradas 1, Edifici RDIT, Campus PMT-UPC, 08860 Castelldefels, Spain}
\affiliation{Institute of Cosmology and Gravitation, University of Portsmouth, Dennis Sciama Building, Portsmouth, PO1 3FX, UK}
\author{Satya Gontcho A Gontcho}
\affiliation{Lawrence Berkeley National Laboratory, 1 Cyclotron Road, Berkeley, CA 94720, USA}
\author{Gaston Gutierrez}
\affiliation{Fermi National Accelerator Laboratory, PO Box 500, Batavia, IL 60510, USA}
\author{Klaus Honscheid}
\affiliation{Center for Cosmology and AstroParticle Physics, The Ohio State University, 191 West Woodruff Avenue, Columbus, OH 43210, USA}
\affiliation{Department of Physics, The Ohio State University, 191 West Woodruff Avenue, Columbus, OH 43210, USA}
\affiliation{The Ohio State University, Columbus, 43210 OH, USA}
\author{Mustapha Ishak}
\affiliation{Department of Physics, The University of Texas at Dallas, 800 W. Campbell Rd., Richardson, TX 75080, USA}
\author{Stephanie Juneau}
\affiliation{NSF NOIRLab, 950 N. Cherry Ave., Tucson, AZ 85719, USA}
\author{Robert Kehoe}
\affiliation{Department of Physics, Southern Methodist University, 3215 Daniel Avenue, Dallas, TX 75275, USA}
\author{Theodore Kisner}
\affiliation{Lawrence Berkeley National Laboratory, 1 Cyclotron Road, Berkeley, CA 94720, USA}
\author{Anthony Kremin}
\affiliation{Lawrence Berkeley National Laboratory, 1 Cyclotron Road, Berkeley, CA 94720, USA}
\author{Martin Landriau}
\affiliation{Lawrence Berkeley National Laboratory, 1 Cyclotron Road, Berkeley, CA 94720, USA}
\author{Laurent Le Guillou}
\affiliation{Sorbonne Universit\'{e}, CNRS/IN2P3, Laboratoire de Physique Nucl\'{e}aire et de Hautes Energies (LPNHE), FR-75005 Paris, France}
\author{Michael Levi}
\affiliation{Lawrence Berkeley National Laboratory, 1 Cyclotron Road, Berkeley, CA 94720, USA}
\author{Marc Manera}
\affiliation{Departament de F\'{i}sica, Serra H\'{u}nter, Universitat Aut\`{o}noma de Barcelona, 08193 Bellaterra (Barcelona), Spain}
\affiliation{Institut de F\'{i}sica d'Altes Energies (IFAE), The Barcelona Institute of Science and Technology, Edifici Cn, Campus UAB, 08193, Bellaterra (Barcelona), Spain}
\author{Aaron Meisner}
\affiliation{NSF NOIRLab, 950 N. Cherry Ave., Tucson, AZ 85719, USA}
\author{Ramon Miquel}
\affiliation{Institut de F\'{i}sica d'Altes Energies (IFAE), The Barcelona Institute of Science and Technology, Edifici Cn, Campus UAB, 08193, Bellaterra (Barcelona), Spain}
\affiliation{Instituci\'{o} Catalana de Recerca i Estudis Avan\c{c}ats, Passeig de Llu\'{\i}s Companys, 23, 08010 Barcelona, Spain}
\author{John Moustakas}
\affiliation{Department of Physics and Astronomy, Siena College, 515 Loudon Road, Loudonville, NY 12211, USA}
\author{Nathalie Palanque-Delabrouille}
\affiliation{Lawrence Berkeley National Laboratory, 1 Cyclotron Road, Berkeley, CA 94720, USA}
\affiliation{IRFU, CEA, Universit\'{e} Paris-Saclay, F-91191 Gif-sur-Yvette, France}
\author{Will Percival}
\affiliation{Department of Physics and Astronomy, University of Waterloo, 200 University Ave W, Waterloo, ON N2L 3G1, Canada}
\affiliation{Perimeter Institute for Theoretical Physics, 31 Caroline St. North, Waterloo, ON N2L 2Y5, Canada}
\affiliation{Waterloo Centre for Astrophysics, University of Waterloo, 200 University Ave W, Waterloo, ON N2L 3G1, Canada}
\author{Claire Poppett}
\affiliation{Lawrence Berkeley National Laboratory, 1 Cyclotron Road, Berkeley, CA 94720, USA}
\affiliation{Space Sciences Laboratory, University of California, Berkeley, 7 Gauss Way, Berkeley, CA  94720, USA}
\affiliation{University of California, Berkeley, 110 Sproul Hall \#5800 Berkeley, CA 94720, USA}
\author{Francisco Prada}
\affiliation{Instituto de Astrof\'{i}sica de Andaluc\'{i}a (CSIC), Glorieta de la Astronom\'{i}a, s/n, E-18008 Granada, Spain}
\author{Ignasi P\'erez-R\`afols}
\affiliation{Departament de F\'isica, EEBE, Universitat Polit\`ecnica de Catalunya, c/Eduard Maristany 10, 08930 Barcelona, Spain}
\author{Graziano Rossi}
\affiliation{Department of Physics and Astronomy, Sejong University, 209 Neungdong-ro, Gwangjin-gu, Seoul 05006, Republic of Korea}
\author{Eusebio Sanchez}
\affiliation{Department of Physics, University of Michigan, 450 Church Street, Ann Arbor, MI 48109, USA}
\author{David Schlegel}
\affiliation{Lawrence Berkeley National Laboratory, 1 Cyclotron Road, Berkeley, CA 94720, USA}
\author{Michael Schubnell}
\affiliation{University of Michigan, 500 S. State Street, Ann Arbor, MI 48109, USA}
\affiliation{Department of Physics, University of Michigan, 450 Church Street, Ann Arbor, MI 48109, USA}
\author{Hee-Jong Seo}
\affiliation{Department of Physics \& Astronomy, Ohio University, 139 University Terrace, Athens, OH 45701, USA}
\author{Joseph Harry Silber}
\affiliation{Lawrence Berkeley National Laboratory, 1 Cyclotron Road, Berkeley, CA 94720, USA}
\author{David Sprayberry}
\affiliation{NSF NOIRLab, 950 N. Cherry Ave., Tucson, AZ 85719, USA}
\author{Gregory Tarl\'{e}}
\affiliation{University of Michigan, 500 S. State Street, Ann Arbor, MI 48109, USA}
\author{Benjamin Alan Weaver}
\affiliation{NSF NOIRLab, 950 N. Cherry Ave., Tucson, AZ 85719, USA}
\author{Rongpu Zhou}
\affiliation{Lawrence Berkeley National Laboratory, 1 Cyclotron Road, Berkeley, CA 94720, USA}
\author{Hu Zou}
\affiliation{National Astronomical Observatories, Chinese Academy of Sciences, A20 Datun Rd., Chaoyang District, Beijing, 100012, P.R. China}

\clearpage

\begin{abstract}
We present the \textsc{SpecDis} value added stellar distance catalog accompanying DESI DR1. \textsc{SpecDis} trains a feed-forward Neural Network (NN) with {\it Gaia} parallaxes and gets the distance estimates. 
To build up unbiased training sample, we do not apply selections on parallax error or signal-to-noise (S/N) of the stellar spectra, and instead we incorporate parallax error into the loss function. Moreover, we employ Principal Component Analysis (PCA) to reduce the noise and dimensionality of stellar spectra. Validated by independent external samples of member stars with precise distances from globular clusters (GCs), dwarf galaxies, stellar streams, combined with blue horizontal branch (BHB) stars, we demonstrate that our distance measurements show no significant bias up to 100~kpc, and are much more precise than {\it Gaia} parallax beyond 7~kpc. The median distance uncertainties are 23\%, 19\%, 11\% and 7\% for S/N $<$ 20, 20 $\leq$ S/N$<$ 60, 60 $\leq$ S/N $<$ 100 and S/N $\geq$ 100. Selecting stars with $\log g<3.8$ and distance uncertainties smaller than 25\%, we have more than 74,000 giant candidates within 50~kpc to the Galactic center and 1,500 candidates beyond this distance. Additionally, we develop a Gaussian mixture model to identify unresolvable equal-mass binaries by modeling the discrepancy between the NN-predicted and the geometric absolute magnitudes from {\it Gaia} parallaxes and identify 120,000 equal-mass binary candidates. Our final catalog provides distances and distance uncertainties for $>$ 4 million stars, offering a valuable resource for Galactic astronomy. 
\end{abstract}


\section{Introduction}
\label{sec:intro}

Our Milky Way (MW) Galaxy is an ideal laboratory to test the physics of galaxy formation and the underlying cosmology, because individual stars in the MW can be resolved by the observers. Various types of information can be extracted from the phase-space distribution of individual stars in the Galaxy, and from its surrounding GCs, satellite galaxies and stellar streams, which enables us to probe the spatial and kinematical structures of the MW disk, bulge, and halo, to infer the assembly history of the MW (galactic archaeology), and construct dynamical models to constrain the nature of the host dark matter halo of our MW. In particular, full 6-dimensional phase-space information, without missing dimensions, is critical for scientific inferences. 

Among the 6-dimensional phase-space information, distances to individual stars can be measured in several different ways. The approaches include, for example, measuring the parallax of individual stars, the usage of the period luminosity relation for RR Lyrae and Cepheids \citep[e.g.][]{2017ApJ...850...96H,2019A&A...622A..60C}, photometric distances to infer the luminosity from stellar color \citep[e.g.][]{2008ApJ...673..864J,2008ApJ...684..287I,2023ApJ...957...65H}, and spectrophotometric distance when stellar atmospheric parameters and distances can be measured together from stellar spectra and photometry \citep[e.g.][]{2014ApJ...784..170X,2019AJ....158..147H,2019ApJ...879...69T,2019ApJS..245...34X,2021ApJS..253...22X,2021ApJ...907...57G,2022ApJS..259...51W,2022A&A...662A..66X, GaiaXp,2024ApJS..273...19Z}. Among all the different methods, the geometric parallax measurements have been revolutionized by {\it Gaia} \citep{2016A&A...595A...1G,Bailer-Jones_2021,2023AJ....166..269B}, but {\it Gaia} does not deliver precise parallaxes at the faint end ($G>17$) \citep{zero_point_correction}, out to large distances ($>\sim5$~kpc) or in crowded regions, and the errors also correlate with proper motion errors. Distances measured for variable stars are relatively accurate, but are not applicable to main sequence stars or giants which contribute the majority of stars. The photometric distance measurements of inferring the luminosity from stellar color are known to suffer from errors of several tens of percent, and may be challenging for hot stars with less color variation beyond the Rayleigh-Jeans tail.
Spectrophotometric distance measurements combine spectroscopy and photometry, with the stellar spectra containing far more information than pure photometric colors, which can provide more precise predictions of the luminosity and distance. The quoted errors in the literature range from about $\sim$10\% to 20-30\%, depending on the spectral type of stars, the signal-to-noise ratio and resolution of the spectrum, and the methodology. 

There are several different ways to perform the spectrophotometric distance measurements. One method involves deriving stellar parameters, and the absolute magnitudes (hence distances) can be inferred through, for example, matching to stellar evolutionary models (stellar isochrones) or calibrated relations based on stars with precisely known distances such as member stars in GCs or with precise parallax measurements \citep[e.g.][]{2010MNRAS.407..339B,2011A&A...532A.113B,2014MNRAS.437..351B,2014ApJ...784..170X,2015AJ....150....4C,2016A&A...585A..42S,2016MNRAS.456..672W,2018MNRAS.481.2970C,2019MNRAS.484..294D,starhorse1,2020A&A...638A..76Q,starhorse2,2025AJ....169..266Y} with probabilistic models. To derive the stellar parameters, physical spectral models are fit to the observed data spectra or apparent magnitudes in different bands. 

Machine learning or deep learning approaches \citep{deep-learning} are extensively utilized to derive stellar parameters and absolute magnitudes, hence distances, from median-to-low-resolution spectra in a data-driven manner \citep[e.g.][]{2017MNRAS.464.3657X,2019AJ....158..147H,2021ApJS..253...22X,2022ApJS..259...51W}. These methods predict the stellar parameters and absolute magnitudes from the stellar spectra and photometry, by learning a model using a sample of stars with more accurate distance and stellar parameter measurements from, e.g., precise {\it Gaia} parallax and other independent, higher-resolution surveys with more accurate stellar parameter estimates.

More recently, many forward modeling methods implemented with machine learning have been developed, such as \cite{2021ApJ...907...57G}, \cite{GaiaXp} and the Payne \citep{2019ApJ...879...69T,2019ApJS..245...34X,2022A&A...662A..66X}. These studies adopt machine learning to build forward models that predict the stellar spectra or apparent magnitude in different bands from a large number of stellar parameters, trained on physical stellar atmospheric models or more precise measurements based on higher resolution spectroscopic survey data. The differentiable stellar spectra are fit to the observed stellar spectra to constrain the stellar parameters, with absolute magnitudes and distances derived as byproducts. 
Especially, Payne has recently been applied by \cite{2024ApJS..273...19Z} to the early survey data release of the Dark Energy Spectroscopic Instrument (DESI). 

DESI is one of the foremost multi-object spectrographs for wide-field surveys \citep{Snowmass2013.Levi,desiScience,desiInstrument,silber22a,SurveyOps.Schlafly.2023,DESI2023a.KP1.SV}. The main science goal of DESI is to achieve the most precise constraint on the expansion history of the Universe to date with baryon acoustic oscillations (BAO) and
other methods \citep{Snowmass2013.Levi}. For the goal, DESI
observes four classes of galactic targets including the
Bright Galaxy Survey targets \citep[BGS;][]{2022APS..APRH13003H},
luminous red galaxies \citep[LRG;][]{2023AJ....165...58Z}, emission line galaxies \citep[ELG;][]{2023AJ....165..126R} and quasars \citep{2023ApJ...944..107C}.

So far DESI has achieved many important science results \citep{desi-collaboration22a}, including BAO signal from galaxies and Lyman alpha forest \citep{2024arXiv240403000D,2024arXiv240403001D}, full-shape galaxy clustering \citep{DESI2024.V.KP5}, two-point clustering statistics \citep{DESI2024.II.KP3} and cosmological constraints \citep{2024arXiv240403002D,DESI2024.VII.KP7B} \footnote{\cite{desi-collaboration22a,DESI2023a.KP1.SV,DESI2023b.KP1.EDR,DESI2024.II.KP3,2024arXiv240403000D,2024arXiv240403001D,DESI2024.V.KP5,2024arXiv240403002D,DESI2024.VII.KP7B} are DESI Collaboration Key Papers.}

In addition to extra-galactic observations, DESI performs one of the so far largest and deepest surveys for stars within our Galaxy. A great number of stellar spectra are being cumulated to understand our own Galaxy and its dark matter halo through stellar kinematics. The Year-1 data has more than 4 million stellar spectra. Accompanying the Year-1 observation, we adopt a data-driven approach to measure the distances of stars from the Year-1 data of the DESI MW Survey \citep[MWS;][]{2023ApJ...947...37C} in this paper. Our attempt is among the various DESI MWS approaches aimed at providing distance measurements for main sequence and giant stars observed by DESI. Our stellar distance catalog (\textsc{SpecDis}), which stands for spectral distances, is published as one of the Value Added catalogs (VAC) of DESI DR1. However, we do not present stellar parameter measurements. The stellar parameter $+$ radial velocity VAC for DESI MWS is measured by DESI MWS RVSpecFit (\textsc{rvs}) pipeline and FERRE spectroscopic pipeline (\textsc{sp}), which are available at \url{https://data.desi.lbl.gov/doc/releases/dr1/vac/mws/} \citep{koposov2025desidatarelease1}.

In this study, we use a feed-forward multilayer perceptron neural network to predict the reciprocal of the square root of luminosity, that is then used to predict the distance of each star. Our validation of the test sample indicates that the precision of distance measurements is approximately 10\% for the high signal-to-noise subset (S/N$>$100). 
Furthermore, we conduct an external validation of our distance estimates using member stars from GCs, dwarf galaxies, the Sagittarius stellar stream (Sgr) and BHB stars, which have precise distance measurements from other methods and can thus serve as a reference. 
This external validation process shows that our distance measurements exhibit no significant bias up to 100~kpc. Based on our distance measurements, we have also proposed a model to distinguish equal-mass binary stars, and have identified a population of 120,000 candidate equal-mass binary systems.

This paper is organized as follows: Section \ref{sec:data} gives an overview of the data from DESI MWS, including the construction of the total sample, the external validation sample and our initial processing of the stellar spectra. Section \ref{sec:Method} introduces the methodology of this work. The main results of our measured distances, precision validation and comparisons with previous studies are presented in Section~\ref{sec:results}. Section \ref{sec:Binary identify} offers further discussion on equal-mass binary star identification. We summarize our findings in Section \ref{sec:concl}.

\section{Data}
\label{sec:data}

\subsection{The DESI Milky Way Survey}

DESI is one of the foremost multi-object spectrograph operating at Kitt Peak National Observatory \citep{Snowmass2013.Levi,desiScience,desiInstrument,desi-collaboration22a,DESI2023b.KP1.EDR}. It features 5,000 fibers, that connect to 3-arm spectrographs ($B$, $R$ and $Z$-arms), altogether span the rest-frame wavelength range of 3,600-9,824~\AA~ with spectral resolution $R\sim2500-5000$ \citep{Corrector.Miller.2023,FiberSystem.Poppett.2024}.

The DESI MW Survey (MWS) is mainly performed at bright time \citep{2023ApJ...947...37C,2024MNRAS.533.1012K}, when the sky conditions are bright due to the moon phase, twilight, and worse seeing conditions making the observation of high redshift galaxies inefficient. The main MWS will observe approximately 7 million stars selected from the DESI Legacy imaging Survey \citep{2019AJ....157..168D} with apparent magnitudes between $r=16$ and $r=19$. It covers most of the Northern Galactic cap region and a significant fraction of the Southern Cap region. DESI MWS also has a backup program \citep{dey2025backupprogramdarkenergy}, that is performed when the observing conditions are poorer with, e.g., very cloudy sky or too bright sky due to clouds, but with no threat of rain and the dome and telescope are usable for observations. It observes the spectra for several million stars mostly brighter than those observed in the main survey. The source selection of MWS backup program is based on {\it Gaia} photometry and astrometry, which does not rely on the photometry of the DESI Legacy imaging Survey, and it involves more stars at lower Galactic latitudes. In this study, we include stars from both the bright time and from the backup observations.

Following the Early Data Release \citep[EDR;][]{DESI2023b.KP1.EDR} of DESI in June 2023, \cite{2024MNRAS.533.1012K} published a DESI MWS stellar value-added catalog for 400,000 stars, including commissioning and science validation data. The first data release (DR1) covering DESI Year-1 observation (observed from May 2021 through June 2022) will soon be made in the spring of 2025 (\cite{desi_yr1}). In this paper, we use the MWS data from DR1. Our stellar distance catalog (\textsc{SpecDis}) is published as one of the Value Added catalogs (VAC) accompanying DESI DR1.

\subsection{Data reduction and pipeline}

DESI stellar spectra are processed first by the general Redrock spectral fitting code \citep{Spectro.Pipeline.Guy.2023}, followed by the pipelines specialized for stellar spectra (\textsc{rvs}, \textsc{sp} and \textsc{wd}\footnote{Here the \textsc{wd} pipeline is developed to process white dwarf spectra.}). Our analysis in this paper depends on \textsc{rvs} and \textsc{sp}. 

Here we provide a brief introduction to \textsc{rvs} and \textsc{sp}. Further details about all of the MWS pipelines are available in \cite{2023ApJ...947...37C}. \textsc{rvs}\footnote{https://github.com/segasai/rvspecfit} derives the radial velocities and atmospheric parameters using the algorithm of \cite{2011ApJ...736..146K}. Interpolated stellar templates from the \textsc{phoenix} model \citep{2013A&A...553A...6H} are fit to the observed stellar spectra (without flux calibration). 

By optimizing the log-likelihood combined across three arms simultaneously, it constrains the radial velocities, effective temperature ($T_\mathrm{eff}$), surface gravity ($\log g$), iron abundance ($[\mathrm{Fe/H}]$), alpha-element abundance ratio ($[\alpha/\mathrm{Fe}]$), the projected stellar rotations and their corresponding uncertainties. 

\textsc{sp} determines stellar atmospheric parameters by inferring individual elemental abundances. It is based on the FORTRAN code, \textsc{ferre}\footnote{https://github.com/callendeprieto/ferre} \citep{2006ApJ...636..804A}, with a new Python package written specifically for MWS, \textsc{piferre}\footnote{https://github.com/callendeprieto/piferre}. Only stellar spectra which have been successfully processed by \textsc{rvs} will be further passed to \textsc{sp}. 
\textsc{sp} stores the continuum normalized data spectra in the rest frame, the best-fitting model spectra with and without continuum normalizations. 

In this paper we only measure the stellar distances, and we will use the stellar parameters measured by the \textsc{rvs} pipeline, which we call as \textsc{rvs} tables, to analyze the quality and statistical properties of our VAC (see Section~\ref{sec:results}). The \textsc{rvs} tables also provide information about spectroscopic fiber information and the cross matched {\it Gaia} DR3 information. We use the continuum normalized spectra in rest frame produced by the \textsc{sp} pipeline.

\subsection{Training and test stellar samples}
\label{subsec:training set}
\begin{figure*} 
\begin{center}
\includegraphics[width=0.9\textwidth]{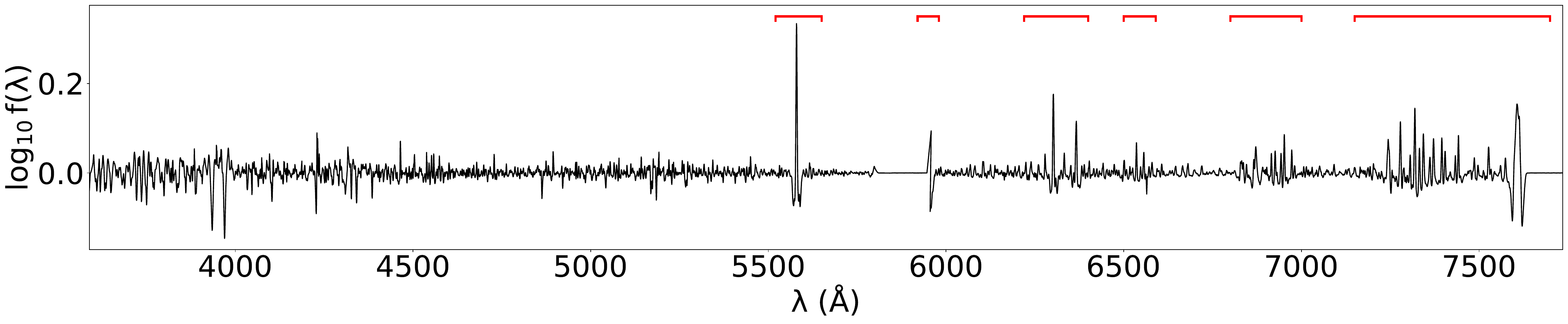}%
\end{center}
\caption{An example of coadded error spectra in log ($y$-axis), obtained by stacking 50,000 individual error spectra. Here $\mathrm{f(\lambda)}$ denotes the flux of error spectra after moving average. The regions of sky lines, marked by red brackets, are discarded. The range devoid of skylines are utilized for PCA. }
\label{fig:skyline}
\end{figure*}

Most studies apply relative error cuts to parallax for their training sample \citep[e.g.][]{2021ApJS..253...22X,2022ApJS..259...51W,GaiaXp}. 
However, such a selection is not ideal for this work based on DESI, because DESI MWS mainly observes stars with $16<r<19$ at bright time. The bright end cut of $r>16$ has eliminated a lot nearby bright giant stars\footnote{DESI backup program observes more giant stars, but the number of nearby giant stars in DESI Year-1 data is still limited.}. Further adopting a relative error cut to parallax would limit the training sample to only those nearby stars, and thus the sample of giants in Year-1 data of DESI MWS used for training would not only be very small, but also subject to biased stellar property distributions\footnote{The training sample selected with $\omega/\sigma_\omega>10$ are dominated by main-sequence stars, and stars with distances greater than 20~kpc in our measurements show different distributions in $T_\mathrm{eff}$-$\log g$ and $\log g$-[Fe/H] space than those in this training sample.} compared with more distant giant stars waiting for distance measurements. Moreover, since DESI is deeper than many previous surveys, those fainter distant halo giants may make this issue more prominent.

S/N cuts to the stellar spectra are often adopted as well \citep[e.g.][]{2021ApJS..253...22X,2022ApJS..259...51W}. However, more distant giants in main DESI MWS have fainter apparent magnitudes hence lower S/N. Adopting S/N selections to the training sample based on Year-1 data of DESI MWS also results in a very small number of giants.

To avoid such selection biases, in this paper we choose not to apply any cuts in parallax error or the S/N of our stellar spectra. This would, however, result in a training sample with significantly less precise parallax measurements and lower quality stellar spectra. To overcome the lowered precision in parallax, we include the error in parallax measurements in the loss function adopted to optimize the network, i.e., stars with more precise {\it Gaia} parallax will be given a higher weight in loss function and vice versa. Moreover, without adopting a selection on S/N, we apply PCA on stellar spectra to reduce the noise and the dimensionality (see Section~\ref{sec:PCA} for more details).

We first select stars with RVS\_WARN$=$0 (from the \textsc{RVS} table) and [Fe/H]$>-$3.9, to ensure they have robust stellar model fits by the \textsc{rvs} pipeline\footnote{RVS\_WARN is a quality control flag of \textsc{rvs}. It is a bit mask flag indicating warnings related to the radial velocity and stellar parameter measurements. The first bit of RVS\_WARN is set to 1 if the discrepancy in the \(\chi^2\) values between the best-fit stellar model and the continuum model is small. The second bit is set to 1 if the radial velocity is $\pm$5~km/s close to the predefined velocity boundary ($-$1500 to 1500~km/s). The third bit is set to 1 if the radial velocity uncertainty surpasses \(100~\text{km/s}\). A spectrum that does not exhibit any of these concerns is indicated by a RVS\_WARN value of zero.} 
and avoid extremely metal-poor stars. These selections remove $\sim$70,000 stars. Moreover, we only use stars with {\it Gaia} RUWE\footnote{RUWE stands for Renormalised Unit Weight Error. For stars with a significantly greater than 1.0 value of RUWE, it maybe a non-single star or has problematic astrometric solution. Detailed definition about RUWE can be found in \url{http://www.rssd.esa.int/doc_fetch.php?id=3757412}.} smaller than 1.2 and {\it Gaia} PHOT\_VARIABLE\_FLAG $!=$ VARIABLE, which can help to eliminate some possible binary stars (see Section~\ref{sec:Binary identify} below) and variable stars. $\sim$ 400,000 stars are excluded with the selections. In addition, we retain only stars with E(B-V)$<$0.5 to avoid stars with significant extinctions. Here only 3,000 stars are excluded. After these selections, we end up with 4 million unique stars without duplicates. We then divide these stars into the training and test samples. The training sample contains a 80\% random subset of the total sample, which includes $\sim$3.2 million stars, and the remaining 20\% test sample contains $\sim$0.8 million stars.

Because sources in the DESI backup observation are selected from {\it Gaia} instead of from the DESI Legacy imaging Survey \citep{2019AJ....157..168D}, and we have included backup sources in this paper, we will utilize {\it Gaia} $G$-band flux throughout this paper, with extinction corrections. As the DESI footprint is at high Galactic latitudes, the 2-dimensional dust map is good enough for dust correction. In this work, we utilize the \cite{dustmap} dust map. We get the extinction coefficient for {\it Gaia} $G$ from \cite{Sartoretti_gaia_extinction_law}. In this paper we use {\it Gaia} DR3 parallaxes for training, with zero-point corrections applied following \cite{zero_point_correction}.

\subsection{Stellar spectra}
\label{sec:stellar_spectra}

We start from the continuum normalized stellar spectra in the rest frame processed by \textsc{sp} which provides DESI bad pixel masks.
We first remove the pixel if it is a bad pixel for more than 10~\% of the stars, and $\sim$50 pixels are discarded. For the other bad pixels, we replace them with the value through linearly interpolating neighboring pixels. We discard the entire $Z$-arm spectra because it is severely contaminated by sky lines. Moreover, we create specific windows to mask pixels contaminated by sky lines in the $B$ and $R$-arms. Figure~\ref{fig:skyline} shows the typical error spectra in log by stacking the errors of 50,000 randomly selected stars. The red brackets in Figure~\ref{fig:skyline} show the masked windows contaminated by sky lines. Finally, the red end of $B$-arm and the blue end of $R$-arm have a narrow overlapping region. We connect $B$ and $R$-arm data by taking the pixel reads from either arm to the middle point of this overlapping region. We end up with 5,500 pixels for each spectrum.

Ultimately, to more efficiently extract information from the stellar spectra, we perform some simple yet crucial transformations. Firstly, because the absorption lines contain the most useful information, we subtract the mean from the continuum normalized spectra, after masking bad pixels and sky lines. This increases the importance of pixels associated with absorption lines. Then we apply the following logarithmic transformation to the spectra. 

\begin{equation}
\label{equ:f-of-x}
S_k = 
\begin{cases} 
\log(L_k+1) & \text{if } L_k > 0, \\
-\log(-L_k+1) & \text{if } L_k \leq 0,
\end{cases}
\end{equation}
here $L_k$ is the value of the $k$-th pixel after subtracting the mean, and $S_k$ is the value of the $k$-th pixel after logarithmic transformation. Because we have subtracted the mean, $L_k$ can be negative. We take the negative sign of $L_k$ when it is negative, and we add unity to make sure that the quantity is positive before the logarithmic transformation. We have verified that the transformation can lead to better precision in our results, as taking the log operations can help us eliminate some extreme pixel values due to noise and make the training more easily to converge. Lastly, we recognize that the normalization of each stellar spectrum is not a constant, and we choose to normalize each stellar spectrum, $S(\lambda)=\{S_k\}$, by $\sqrt{\Sigma_k S_k^2}$ before doing PCA (see Section~\ref{sec:PCA} below).
\begin{figure*} 
\begin{center}
\includegraphics[width=0.9\textwidth]{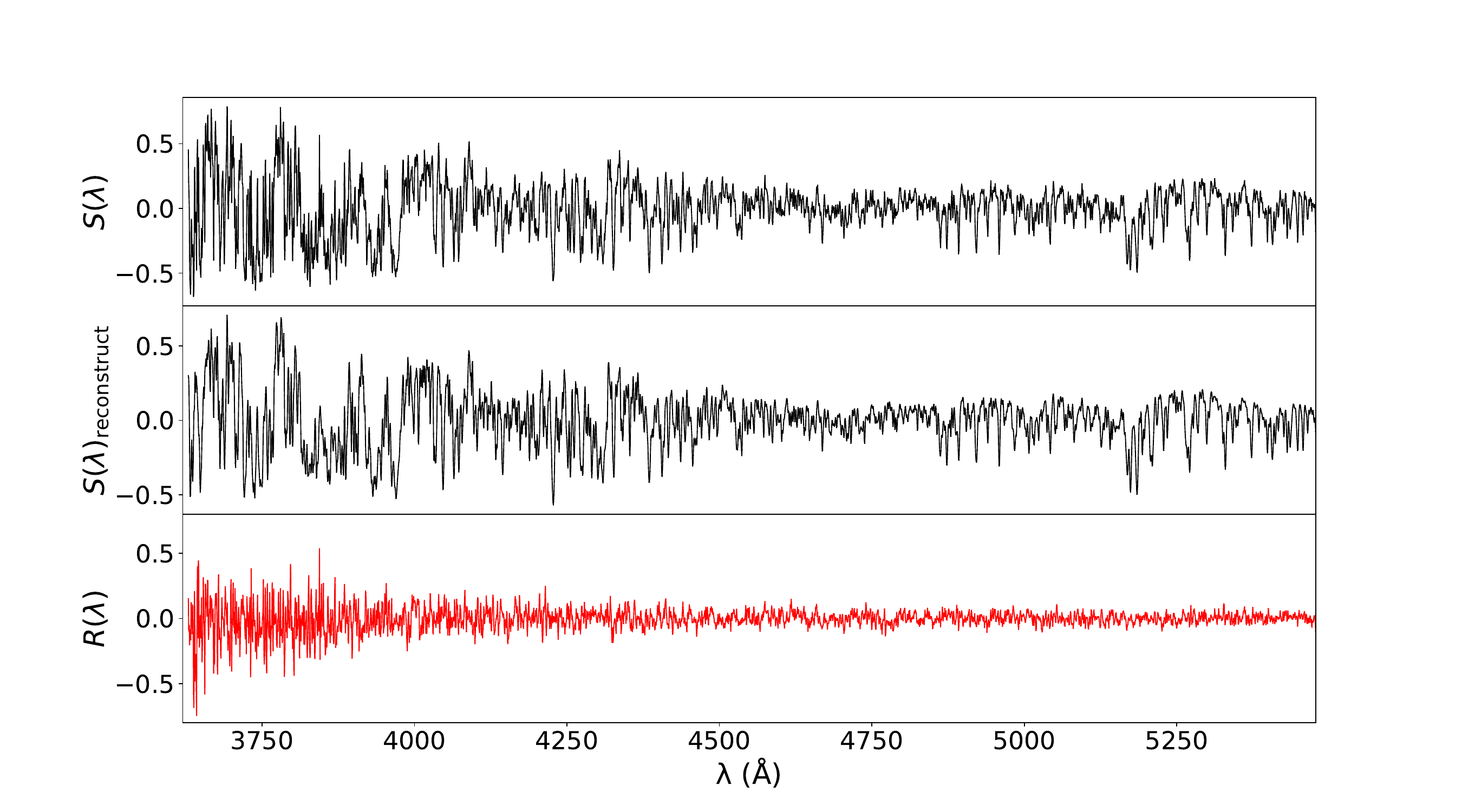}%
\end{center}
\caption{An example of PCA reconstruction, focusing on a DESI B-arm spectrum with a S/N of 3 for a randomly selected star. Here $\mathrm{S(\lambda)}$ in the top panel denotes the log flux of stellar spectra after subtracting the mean and after logarithmic transformation (Equation~\ref{equ:f-of-x} and Section~\ref{sec:stellar_spectra}). The middle panel shows the reconstructed spectrum based on the first 100 principal components. The bottom panel shows the residuals.}
\label{fig:reconstrucion_sn9}
\end{figure*}

\subsection{External validation sample}
\label{sec:validation set}
In this section we will introduce the external validation sample of this work. Because {\it Gaia} parallaxes have very large uncertainties beyond 7~kpc, we require some independent distance measurements as reference to verify our measurements.
We use member stars from GCs, dwarf galaxies, Sgr and BHB stars from DESI Year-1 data with accurately known distances, as our external validation sample. In this paper, we will denote the accurate distance modulus inferred from these member stars by $(m-M)_\mathrm{true}$. 

We cross match our sample of stars (Section~\ref{subsec:training set}) with the member star catalogs of GCs \citep{2021MNRAS.505.5957B} and dwarf galaxies \citep{2022ApJ...940..136P} according to their coordinates. The typical relative distance uncertainty for GCs is smaller than 1\%, and for dwarf galaxies it is $\sim$5\%. For the matched stars, we further utilize their radial velocities (RVs) in DESI to select secure member stars. We first calculate the average and 1-$\sigma$ scatter in their RVs. We maintain only those stars whose RVs differ by less than 2-$\sigma$ from the average RVs for validations. 

The member stars from Sgr are from \cite{2021MNRAS.501.2279V}, with their distances calibrated using RR-Lyrae stars and typical relative distance errors are $\sim$10\%. We cross match Sgr catalog with our stellar sample by {\it Gaia} source ID.
For BHB stars, we cross match with our stellar sample by DESI TARGETID, i.e., DESI source ID. 
Details about the BHB sample in DESI can be found in \cite{2024arXiv241009149B}.

Ultimately, we have 2,154 member stars from GCs (mainly covering 5 to 20~kpc), 320 member stars from dwarf galaxies (covering 80-100~kpc, and several stars could approach 200~kpc), 1,001 member stars from Sgr (25-60~kpc) and 4,326 stars from DESI Year-1 BHB sample (8-100~kpc). Nearby GC member stars can include both main sequence stars and giants, while member stars from Sgr and more distant dwarf galaxies are all giants. 

\section{Methodology}
\label{sec:Method}
In this section we introduce the application of PCA to the DESI stellar spectra, the neural network methodology and the error model for the \textsc{SpecDis} VAC.

\subsection{PCA to reduce the noise and dimensionality in the data spectrum}
\label{sec:PCA}

In this work, we adopt PCA on stellar spectra to reduce the noise and the dimensionality of the data set. This follows \cite{2021ApJS..253...22X}. PCA is a standard multivariate analysis technique that is frequently used in many different fields of astronomy and astrophysics \citep[e.g.][]{2009MNRAS.394.1496B,2012MNRAS.421..314C,2023MNRAS.523.5789Z}. The concept is that a spectrum with $N$ pixels can be regarded as a single point in an $N$-dimensional space. A group of spectra forms a cloud of points in this high-dimensional space. PCA searches for $N$ vectors, known as principal components (PCs), which have decreasing orders of variance in the cloud of points. Each time a PC with the highest variance is determined, the succeeding PC in turn to be searched should have the highest variance under the constraint that it is orthogonal to or uncorrelated with the preceding components.

After deducing the $N$ vectors or PCs, the $i$-th stellar spectrum, which we define and denote as $S(\lambda)_i=\{S_{i,k}\}$ can be decomposed as,

\begin{equation}
S_{i,k}=\sum_{j}{C_{i,j}E_{j,k}}+R_{i,k},
\label{equ:sumpcs}
\end{equation}
where the index $k$ represents the $k$-th pixel. $E_{j,k}$ represents the $k$-th pixel value for the $j$-th PC, and $C_{i,j}$ is the corresponding coefficient for $j$-th PC. $\sum_{j}{C_{i,j}E_{j,k}}$ gives the reconstructed spectrum for the $i$-th stellar spectrum, and we denote the reconstructed spectrum by $S(\lambda)_{i,\mathrm{reconstruct}}=\{S_{i,k,\mathrm{reconstruct}}\}$. $R(\lambda)_i=\{R_{i,k}\}$ is the associated residual.

Figure~\ref{fig:reconstrucion_sn9} shows an example of reconstruction using the first 100 PCs. The S/N of the original spectrum is only 3. The reconstructed spectrum using the first 100 PCs is shown in the middle panel, which is less noisy. The bottom panel shows the residual, which is dominated by noise. It is straight-forward to understand the reduction of noise by using the reconstructed spectra from several top PCs, if their largest variances correspond to the most useful information in the stellar spectrum. On the other hand, we expect the remaining less important PCs to be dominated more by noise. Moreover, based on the key concept of PCA, using the first 100 coefficients, $C_{i,1-100}$, is equivalent to using the reconstructed spectrum from the first 100 PCs, and thus we use the first 100 coefficients for each star for training, instead of the spectrum, throughout this paper. The approach naturally helps us to reduce the dimensionality in the data cube. 

Figure~\ref{fig:pc amplitude} in Appendix A presents the correlations between the [Fe/H], $T_\mathrm{eff}$ and the coefficients of the top two PCs and the 100-th PC. The coefficients of the top PCs are correlated with these physical features, and the coefficient of the 100-th PC shows less correlation. We have also confirmed that higher-order PCs show almost no correlations with the stellar features.

\subsection{The Neural Network Models}
\subsubsection{Training Label and Loss Function}
\label{sec:label_loss}

Neural network (NN) establishes the connection between the input stellar spectra and the quantity to be predicted, and the predicted quantity is called label. To establish this connection, usually an appropriate loss function is defined, that is going to be minimized to achieve the best trained model. In this section, we introduce our choice of the label and the loss function.

In this paper we utilize the reciprocal of the square root of luminosity (Equation \ref{equ:label}) as our label, i.e., the NN will predict the label defined by Equation~\ref{equ:label} below for each star 

\begin{equation}
\mathrm{label}\equiv\frac{1}{\sqrt{\it L_{G}}}\equiv\frac{\omega}{10^{2-m_{G}/5}}\equiv\frac{\omega}{\sqrt{F_{G}}}.
\label{equ:label}
\end{equation}

Here $\omega$ is the {\it Gaia} parallax. $L_{G}$ is the $G$-band luminosity. The deduction is based on the relation between absolute magnitude ($M_G$) and apparent magnitude ($m_G$) of $m_G-M_G=-5\log_{10}\frac{\omega}{100}$, where $\omega$ is in units of mas, and $M_G=-2.5\log_{10}L_G$. $F_{G}$ is the observed flux in {\it Gaia} $G$ band, and is related to the apparent magnitude as $\sqrt{F_{G}}=10^{2-m_{G}/5}$. Here we ignore the error of observed flux, which is very small compared with the parallax error. Equation~\ref{equ:label} is linear in parallax, and thus adopting Equation~\ref{equ:label} as label ensures that the prediction is unbiased even with a large parallax error.

The loss function measures the difference between the predicted label and the true label in an NN model. It quantifies the performance of an NN model. 
Once the NN predicts the label for the $i$-th star, $\mathrm{label}_i$, the loss function is constructed with respect to parallax, as shown by Equation~\ref{equ:label} below: 

\begin{equation}
\chi^2 = \sum \frac{(\mathrm{label}_i \times 10^{2-m_{G,i}/5} - \omega_i)^2}{\sigma_{\omega_i}^2}.
\label{equ:lossfunc}
\end{equation}

Our loss function is a $\chi^2$ statistic. $\omega_i$ and $\sigma_{\omega_i}$ are the {\it Gaia} parallax and parallax error for star $i$, and $m_{G,i}$ is the apparent magnitude. The summation goes over all stars in the training sample. Though we did not include selections on parallax error for our training sample, the loss function in Equation~\ref{equ:lossfunc} naturally assigns smaller weights to stars with larger parallax errors in the training process. 

Ultimately, we obtain the distance or distance modulus by using the predicted label, which provides the luminosity or absolute magnitude, and compare it with the observed apparent magnitude of each star.

\subsubsection{Neural Network Structure}
\label{sec: network structure}
We adopt a feed-forward multilayer perceptron NN model that maps the first 100 coefficients of PCs to the label defined in Equation~\ref{equ:label}. Here feed-forward means all information flows forward only in the network. Multilayer refers to the structure of an NN that is composed of multiple layers, where each layer is connected to the next layer. Adopting the Einstein summation convention, our NN contains four layers as

\begin{equation}
\label{equ:Mks}
\mathrm{label} = w_k^3 I(w_{jk}^2(w_{ij}^1 I(w_{{\lambda}i}^0 f_{\lambda} + b_i^0) + b_{j}^1) + b_{k}^2) + b^3,
\end{equation}
where I is the ReLU\footnote{ReLU is an activation function, which returns 0 when the inputs are smaller than 0 and returns the original values if the inputs are greater than or equal to 0.} activation function, $w$ and $b$ are weights and biases of the network to be optimized, which we call neurons, and $f$ denotes the coefficients of PCs. We adopt 100 neurons for all four layers. The so-called perceptron represents a single neuron-like unit that performs a specific computation.

\subsection{Error Model}
\label{sec:error model}
In this section, we discuss the methodology for quantifying the measurement uncertainty. For simplicity, we will transform the NN-predicted label of $1/\sqrt{L_{G}}$ (see Section~\ref{sec:label_loss}) to the absolute magnitude ($M_{G,\mathrm{NN}}$) for the remainder of this paper. We will present the estimated error statistics in Section~\ref{sec:error estimation}.

\subsubsection{Measurement Uncertainty for each star}
\label{measurement uncertainty}

We provide error estimates for $M_{G,\mathrm{NN}}$ and distances by Monte-Carlo sampling the error spectra of individual stars. Here the error spectra give the associated error for each pixel read of the stellar spectra, which encompass a combination of read-out noise, Poisson errors and errors in sky subtraction by the pipeline. We start with the continuum normalized stellar spectra in the rest frame, to which we add Gaussian noise in accordance with the error spectrum read at each pixel. 

This process is repeated to generate twenty randomly perturbed versions of each spectrum. We then repeat all steps outlined in \ref{sec:stellar_spectra} to process each perturbed spectrum: subtracting the mean, taking the logarithm, and normalizing the spectrum. Subsequently, PCA is utilized to reduce the noise and dimensionality. Notably, the PCs are fixed to be those we obtained in Section~\ref{sec:PCA} based on the data spectra. We did not do PCA again for each set of perturbed spectra, but instead we utilize the same set of PCs for decomposition and getting the coefficients. Finally, we input the coefficients of the top 100 PCs into our trained NN to obtain twenty sets of $M_{G,\mathrm{NN}}$ or distance for each star. 

The standard deviation of these values serves as the measurement uncertainty of $M_{G,\mathrm{NN}}$ or distance for each star. In our analysis, we have chosen to ignore the uncertainties from the apparent magnitudes of each star, as the photometric uncertainties are much smaller than those from the noisy stellar spectra.

\subsubsection{Training uncertainty}
The inherent uncertainty of NN is estimated by training several NNs. Each NN is initialized with a distinct random seed, thereby yielding some different estimates of $M_{G,\mathrm{NN}}$ and distance for each star. However, we do not estimate the uncertainty of each individual star as in Section~\ref{measurement uncertainty}, because the uncertainty is an intrinsic feature of NN model, so it is a fixed value instead of varying among different stars. The uncertainty is the standard deviation of the difference between different NN predicted $M_{G,\mathrm{NN}}$. This uncertainty is typically $\sim$0.05 magnitude. The total uncertainty of a star is the square root of the measurement and training uncertainties added in quadrature.

\begin{figure} 
\begin{center}
\includegraphics[width=0.49\textwidth]{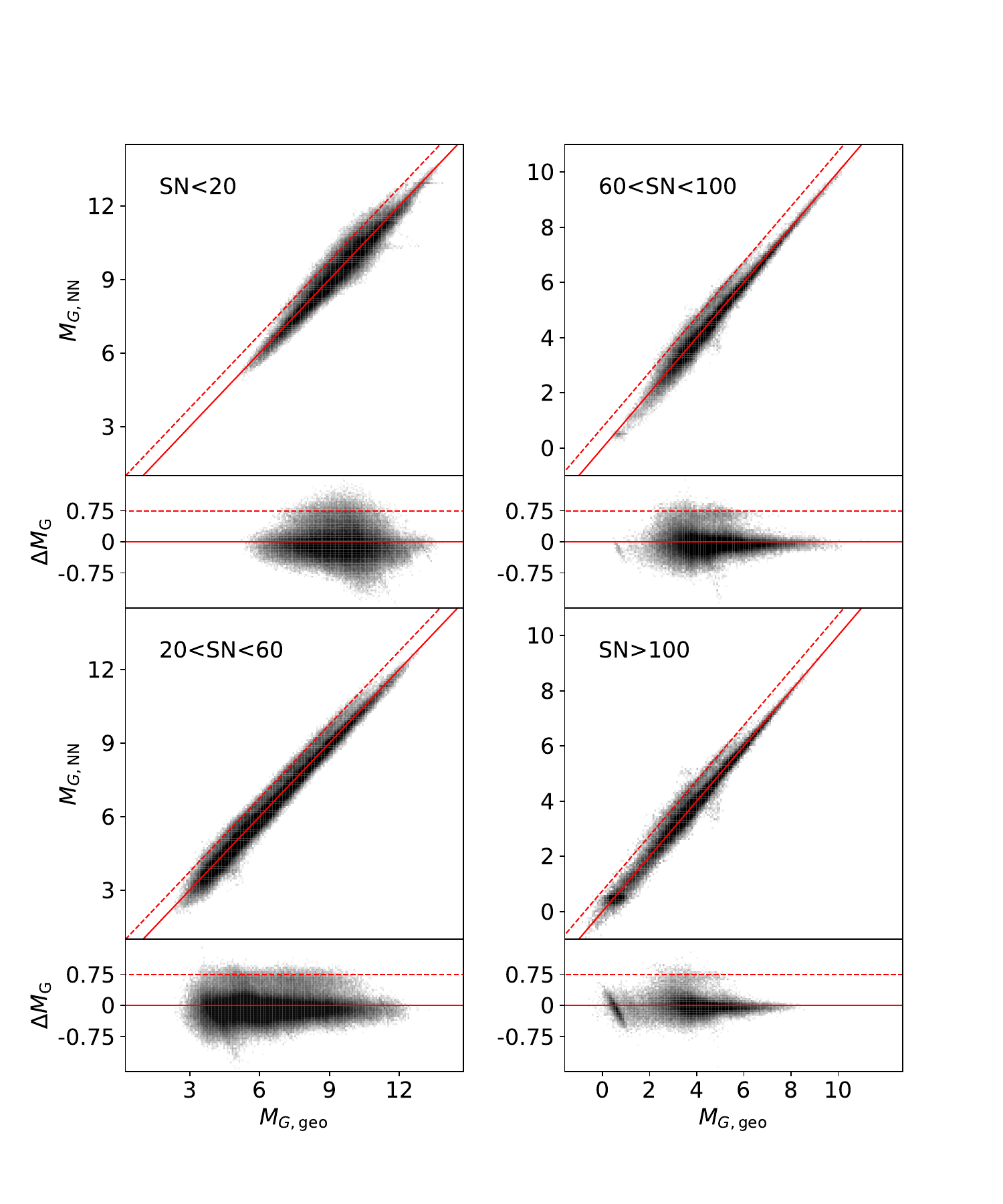}%
\end{center}
\caption{Comparison between $M_{G,\mathrm{NN}}$ and $M_{G,\mathrm{geo}}$. This is based on stars with precise {\it Gaia} parallax measurements ($\omega/\sigma_{\omega}>10$) of the test sample. Four subsamples are shown according to different ranges in S/N of their spectra, with the S/N range shown by the text in corresponding panels. 
The residuals of $\Delta M_{\text{G}}=M_{G,\mathrm{NN}}$ - $M_{G,\mathrm{geo}}$ are also shown at four lower bottom panels. The red solid line marks $M_{G,\mathrm{NN}}$ = $M_{G,\mathrm{geo}}$, and the dashed line is offset by 0.75 magnitude from the red solid line. There are two-band structures: single stars are distributed along the solid line, while the dashed line may correspond to binary systems (further detailed in Section~\ref{sec:Binary identify}).}
\label{fig:sn mag}
\end{figure}

\begin{figure} 
\begin{center}
\includegraphics[width=0.49\textwidth]{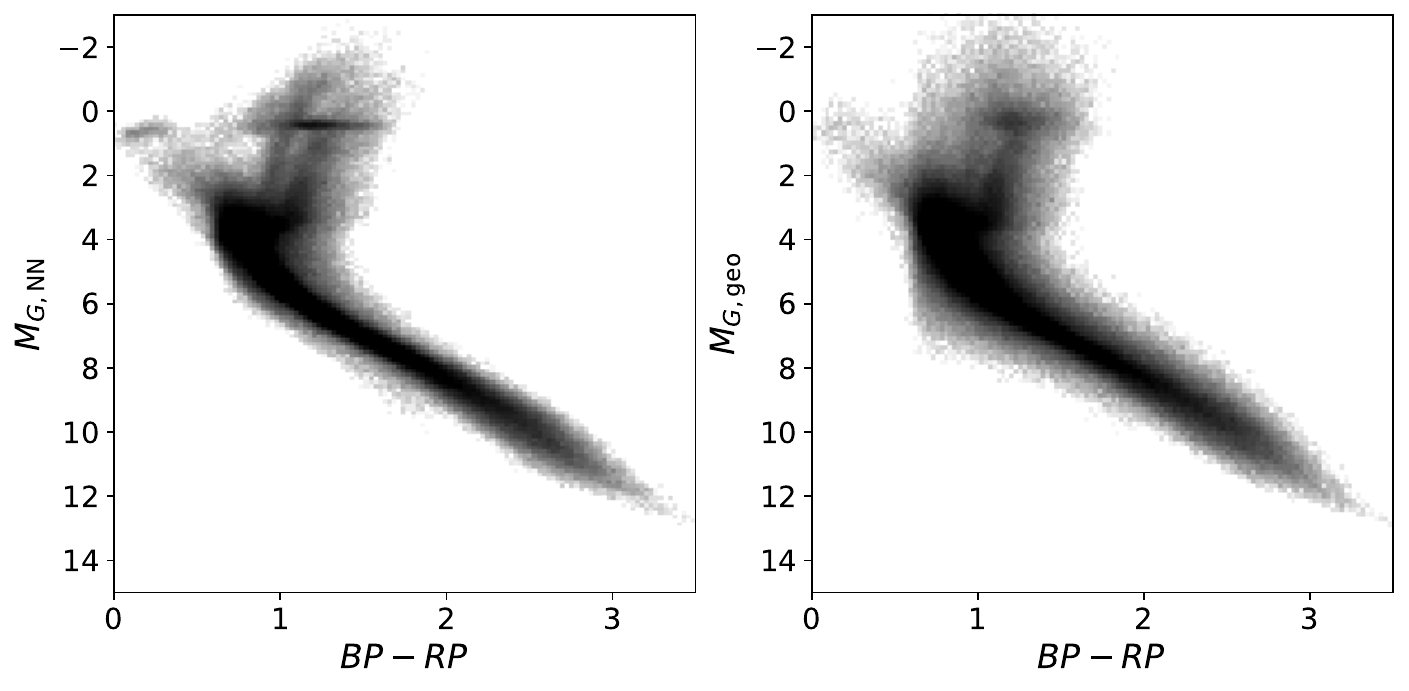}%
\end{center}
\caption{The Hertzsprung-Russell diagram based on the full test sample. The $x$-axis of two panels is the color index derived from the mean magnitudes of the {\it Gaia} photometry in the $BP$ and $RP$ bands. The $y$-axis of the left panel represents the NN predicted absolute magnitude, $M_{G,\mathrm{NN}}$. The $y$-axis of the right panel shows the absolute magnitude deduced from {\it Gaia} parallax, $M_{G,\mathrm{geo}}$, without any cuts on parallax error. Negative parallaxes are excluded from the right plot.}
\label{fig:hr diagram}
\end{figure}

\section{Results}
\label{sec:results}

In this section, we will test and validate the accuracy of our \textsc{SpecDis} VAC, using both the test sample and the independent external validation sample. We will present statistical summaries of our distance catalog for the DESI Year-1 data. Comparisons with previous studies will also be shown.

\subsection{Validation on the test sample}
\label{sec:Results of training and test sets}

Figure~\ref{fig:sn mag} compares the NN predicted absolute magnitude, $M_{G,\mathrm{NN}}$, and the absolute magnitude calculated from {\it Gaia} parallax with zero point corrections \citep{zero_point_correction}, $M_{G,\mathrm{geo}}$. 
To have a robust validation, Figure~\ref{fig:sn mag} is based on a subset of the test sample with precise {\it Gaia} parallaxes of $\sigma_{\omega}/\omega<0.1$, so that $M_{G,\mathrm{geo}}$ can be used as the reference for comparison. We show results of four subsamples with different S/N ranges\footnote{The average S/N of the $B$ and $R$ arms.} (see the text in each panel). We show both $M_{G,\mathrm{NN}}$ and the difference between $M_{G,\mathrm{NN}}$ and $M_{G,\mathrm{geo}}$ ($\Delta M_{G}$, lower panels) as a function of $M_{G,\mathrm{geo}}$. All panels demonstrate high precision for $M_{G,\mathrm{NN}}$ with no particular bias, even in the panel with the lowest S/N. As S/N increases, the scatter of $M_{G,\mathrm{NN}}$ decreases. The 1-$\sigma$ scatters from the lowest to the highest S/N panels are 0.48, 0.26, 0.24 and 0.22 magnitudes, which correspond to distance uncertainties of 24\%, 13\%, 12\%, and 11\%, respectively.

Figure~\ref{fig:sn mag} shows a two-band structure, and we interpret this as the existence of unresolved binaries.
Sources that fall along the red solid diagonal line are single stars, and the red dashed line represents binaries. For binaries, the predicted $M_{G,\mathrm{NN}}$ is fainter than the geometric magnitude $M_{G,\mathrm{geo}}$, because $M_{G,\mathrm{geo}}$ is the combined magnitude from two stars, so this shifts binary stars away from the diagonal line. In all panels, the red dashed line is shifted upwards by 0.75 magnitudes from the red solid line. This corresponds to equal-mass binaries, which have doubled combined flux, so the geometric magnitude of the two stars is expected to be 0.75 magnitudes brighter than that of a single star, whereas $M_{G,\mathrm{NN}}$ remains equivalent to that of a single star because two equal-mass stars have similar spectral types, and thus the combined spectrum remains the same as that for a single star. 

\begin{figure*} 
\begin{center}
\includegraphics[width=0.4\textwidth]{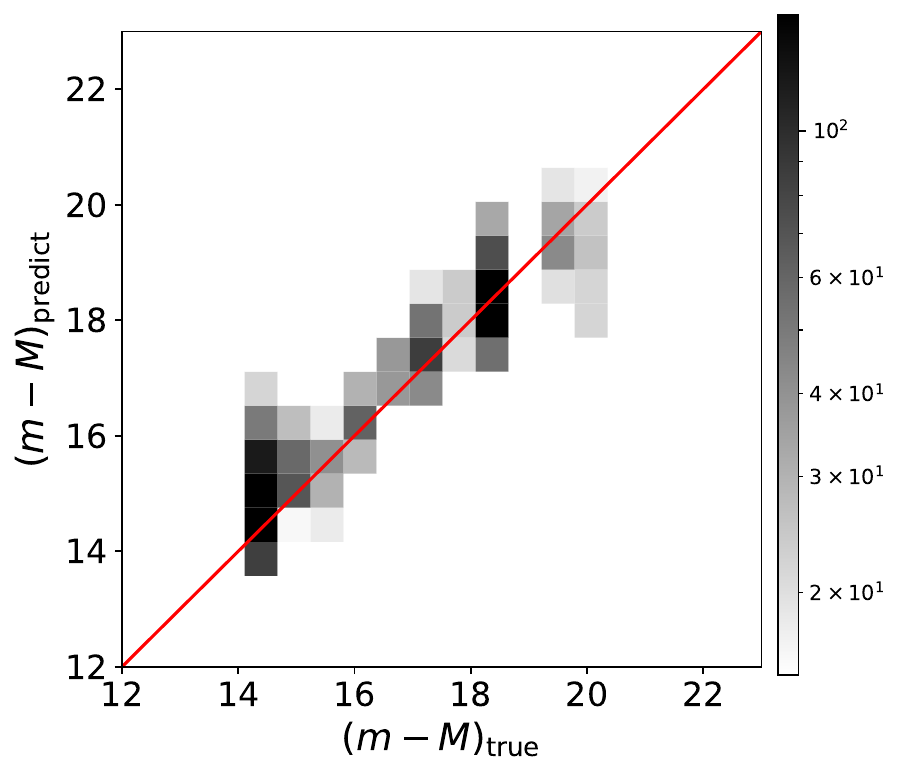}%
\includegraphics[width=0.4\textwidth]{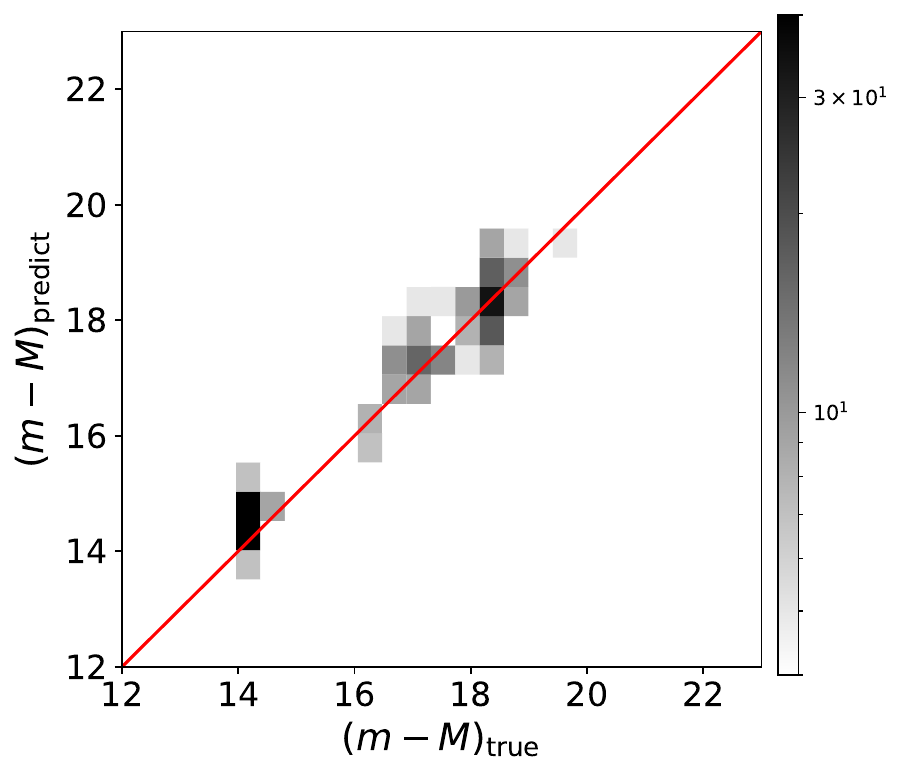}%
\end{center}
\caption{Comparison between NN predicted distance modulus, $(m-M)_\mathrm{predict}$, and reference distance modulus, $(m-M)_\mathrm{true}$. The reference distance modulus is from GCs, Sagittarius stream and dwarf galaxy member stars that can have a more precise distance measurement from their hosts or stream track. {\bf Left:} All validation stars. {\bf Right:} Stars with distance modulus error $<$ 0.5, which correspond to a precision of 25\% in distance.}
\label{fig:validation}
\end{figure*}

\begin{figure} 
\begin{center}
\includegraphics[width=0.4\textwidth]{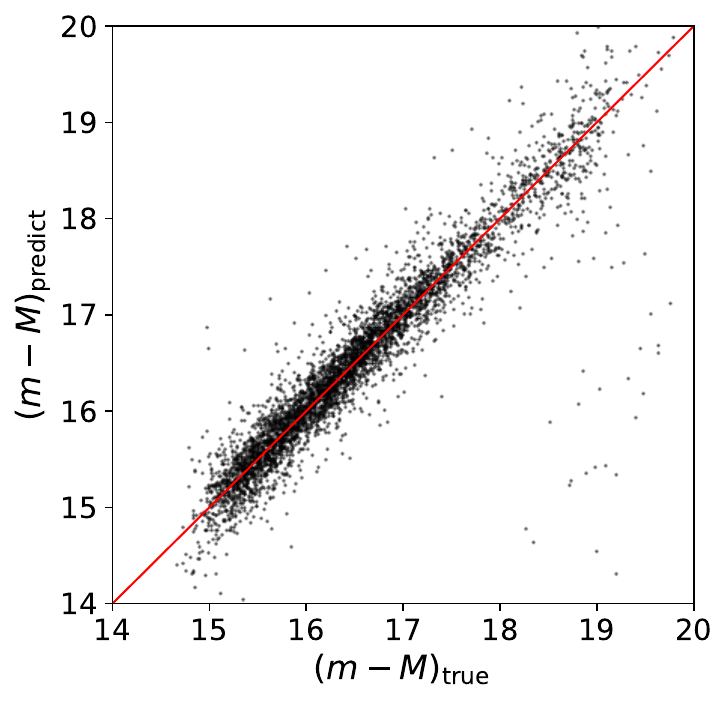}%
\end{center}
\caption{Comparison between $(m-M)_\mathrm{predict}$ and $(m-M)_\mathrm{true}$ based on all DESI Year-1 BHB stars. The reference distance modulus, $(m-M)_\mathrm{true}$, is from DESI Year-1 BHB VAC. }
\label{fig:bhb validation}
\end{figure}

Figure~\ref{fig:hr diagram} shows the distribution of the full test sample in the HR diagram. The $x$-axis represents the color index derived from the mean magnitudes in the {\it Gaia} BP and RP bands after dust correction. And the $y$-axis of the two panels shows the $M_{G,\mathrm{NN}}$ and $M_{G,\mathrm{geo}}$.
The left panel of Figure~\ref{fig:hr diagram} provides a sharper visual representation of the various evolutionary stages of stars than the right panel. 
It presents clear red clump (RC, $1<BP-RP<1.6$) and BHB features ($0<BP-RP<0.3$) when $M_{G,\mathrm{NN}}$ is around 0.5 magnitudes. It also shows a clear dual red giants branch (RGB) feature. This is due to a split in metallicity, with the more metal-poor part belonging to Gaia\protect-–Sausage\protect-–Enceladus \citep[e.g.][GSE]{2018MNRAS.478..611B,2018Natur.563...85H} and more metal-rich one belonging to MW disk. As a comparison, the right plot, which is based on {\it Gaia} parallax without any error cuts, shows more fuzzy BHB, RC and RGB features. The comparison shows that our distance measurements of RGB are more precise than those inferred from {\it Gaia} parallax. This is mainly due to additional information brought in by the DESI stellar spectra. Notably, if we show in the right plot only those stars with precise {\it Gaia} parallax ($\omega/\sigma_{\omega}>10$), the right plot would look as sharp as the left one. However, those stars with precise Gaia parallaxes are mainly those within 5~kpc, whereas the stars in the left plot extends to much larger distances based on the full test sample. 

\subsection{External Validation}
\label{sec:Results on validation sets}

In this section, we validate our measurements by comparing the NN predicted distance modulus, $(m-M)_\mathrm{predict}$, with the reference from the external validation set, $(m-M)_\mathrm{true}$ (see Section~\ref{sec:validation set}). Figure~\ref{fig:validation} shows the comparison between $(m-M)_\mathrm{predict}$ and $(m-M)_\mathrm{true}$. 

The left panel of Figure~\ref{fig:validation} covers a distance modulus range from 14 to 22 magnitudes, corresponding to a distance range from approximately 6~kpc to 250~kpc. The 1-$\sigma$ scatter of $(m-M)_\mathrm{predict}$ is about 1 magnitude, which corresponds to a distance uncertainty of $\sim$50\%. The $(m-M)_\mathrm{true}$ of most stars from GCs is smaller than 16 magnitude, and Sgr member stars span from 16 to 19 magnitude. 
Almost all member stars from dwarf galaxies have $(m-M)_\mathrm{true}$ greater than 19 magnitudes. The distribution of $(m-M)_\mathrm{predict}$ in the left panel is roughly centered around the diagonal line, indicating that our NN model exhibits no significant bias in its distance predictions. 

The right panel of Figure~\ref{fig:validation} shows $(m-M)_\mathrm{predict}$ versus $(m-M)_\mathrm{true}$ after excluding stars with distance modulus or $M_{G,\mathrm{NN}}$ errors greater than 0.5 magnitudes. After excluding these stars, the scatter is significantly reduced as expected, and the prediction of $(m-M)_\mathrm{true}$ shows no bias when compared to the distribution observed in the left panel.

The right panel of Figure~\ref{fig:validation} also demonstrates that our error model (see Section~\ref{sec:error model}) is reliable. As the distance modulus uncertainty calculated from the difference between $(m-M)_\mathrm{predict}$ and $(m-M)_\mathrm{true}$ of the right panel roughly equals the uncertainty threshold of 0.5 magnitudes that we apply, corresponding to distance uncertainty of 25\%. We will provide more details about the measurement error in Section~\ref{sec:error estimation} below.

However, for the most nearby and most distant stars in the left panel of Figure~\ref{fig:validation}, the prediction presents some bias. The distances are overestimated at $(m-M)_\mathrm{true}\sim 14.5$, with $(m-M)_\mathrm{predict}$ brighter than $(m-M)_\mathrm{true}$. 
For a detailed overview of these predictions, see Table~\ref{tab:validation_table} in Appendix~B, which lists the median values of the NN predicted distances for GCs and dwarf galaxies.
Additionally, we show in Figure~\ref{fig:overestimated gc} of Appendix~B the distribution of NN predicted distance modulus and geometric distance modulus from {\it Gaia} parallax for each of the GC, dwarf galaxy and different distance bins of Sgr. The slight bias at $(m-M)_\mathrm{true}\sim$14.5 are from NGC 5904 and NGC 6205.
We are uncertain about the cause behind this, and we have checked that differences among various literature measurements cannot fully account for this bias.
But after excluding stars with distance modulus errors greater than 0.5 magnitudes, which is equivalent to excluding low S/N stars, there is no particular bias in the right panel of Figure~\ref{fig:validation}.

For the most distant stars with $(m-M)_\mathrm{true}$ $>$ 19 in the left panel of Figure~\ref{fig:validation}, $(m-M)_\mathrm{predict}$ is slightly underestimated. All of these stars are giants from dwarf galaxies. In fact, the number of stars with $(m-M)_\mathrm{true}$ $>\sim$ 19 is quite few as shown in Figure~\ref{fig:overestimated gc} of Appendix B, and is mainly from Draco and Sextans. The slight underestimate in distance at $(m-M)_\mathrm{true}$ $>$ 19 is mainly because the number of very distant stars is significantly less than the number of nearby stars, and distant stars also have lower weights in our loss function (see Equation~\ref{equ:lossfunc}). 
If including selections on parallax error and S/N for the training sample, we find distant stars in Figure~\ref{fig:validation} would suffer from more significant biases.

Despite the small amount of bias at the nearest and furthest ends, interestingly, Figure~\ref{fig:overestimated gc} of Appendix B clearly shows that our measured distances perform significantly better than those based on the {\it Gaia} parallaxes beyond 7~kpc. Our distances show much smaller bias and dispersion than the distances from {\it Gaia} parallaxes in NGC 6205 at 7.42~kpc and most of the other more distant systems, when the number of member stars is large enough. This proves the success in our methodology and the robustness of our distance measurements. We utilize {\it Gaia} parallaxes for training, but the predicted distances are much more precise than {\it Gaia} parallaxes out to $\sim$100~kpc as confirmed by this external validation.

Figure~\ref{fig:bhb validation} further presents the comparison between $(m-M)_\mathrm{predict}$ from our NN and $(m-M)_\mathrm{true}$ from DESI Year-1 BHB sample as reference. It covers a wide range of distance modulus from 8 to 20 magnitudes, which corresponds to a distance range from 8 to 100~kpc. The 1-$\sigma$ scatter is only 0.36 magnitude, corresponding to a distance uncertainty of 18\%. For BHBs, there is no bias, and the scatter is smaller than Figure~\ref{fig:validation}. We find that the median S/N of our BHB sample is 34, while the median S/N of member stars in our GCs, dwarf galaxies, and Sgr external validation sample are only 9, 11 and 13 respectively. Thus S/N dominates the precision of our prediction. Lower S/N leads to a larger scatter and slight bias in the left plot of Figure~\ref{fig:validation}.

Figures~\ref{fig:validation}, \ref{fig:bhb validation} and \ref{fig:overestimated gc} of Appendix B demonstrate that our NN predicted distances demonstrate no significant bias up to $\sim$100~kpc. We claim that the distance catalog used in this work is considered credible because of its thorough validation against independent distance measurements, rigorous error analysis, and transparency in the methodologies applied. These factors ensure a high level of confidence in the accuracy of the distances provided.

\begin{figure} 
\begin{center}
\includegraphics[width=0.4\textwidth]{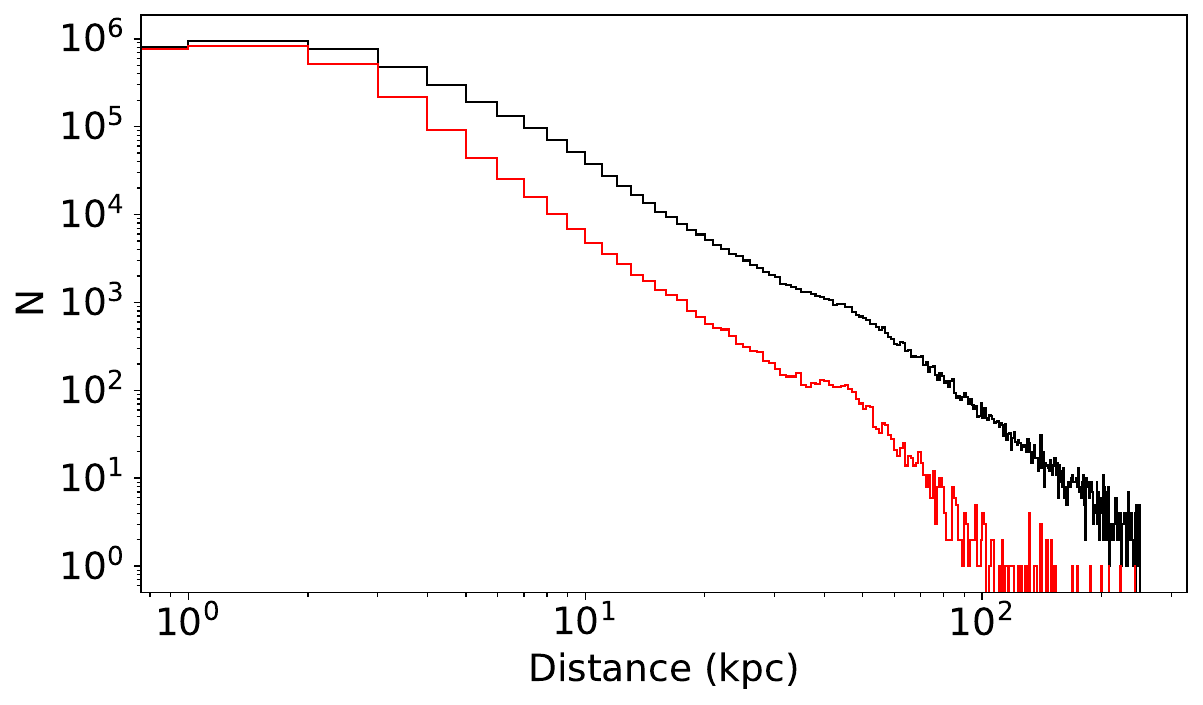}%
\end{center}
\caption{{\bf Black line:} Distribution of heliocentric distance of the full sample. {\bf Red line:} Distribution of heliocentric distances of stars with relative distance error smaller than 25\%.}
\label{fig:distance distribution}
\end{figure}

\subsection{Statistics of the Distance catalog}

\subsubsection{Distance}

Figure~\ref{fig:distance distribution} presents the number of stars as a function of heliocentric distance. The black is for all stars. Our measurements cover a wide range of distance. 96~\% stars are located within 15~kpc. About 120,000 stars have distances greater than 15~kpc,
and about 20,000 are beyond 50~kpc. A number of stars could even lie beyond 200~kpc. The red histogram shows those stars with relative distance error smaller than 25\%. This results in a much smaller sample. 

Both the red and black histograms in Figure~\ref{fig:distance distribution} show a double power-law form, with the break radius at $\sim$50~kpc. The existence of a break radius in the radial density profile of MW stellar halo has been reported in many previous studies \citep[e.g.][]{10.1111/j.1365-2966.2009.15242.x,10.1111/j.1365-2966.2011.19237.x,Hernitschek_2018,Han_stellar_halo_density_profile}. The break radius may indicate a massive accretion event. However, Figure~\ref{fig:distance distribution} here only shows a very preliminary radial distribution of stellar number counts, without correcting for any survey selection functions. We leave more detailed investigations of the halo star density profile to a future study.

\subsubsection{Distance Uncertainty}
\label{sec:error estimation}

\begin{figure}
\begin{center}
\includegraphics[width=0.4\textwidth]{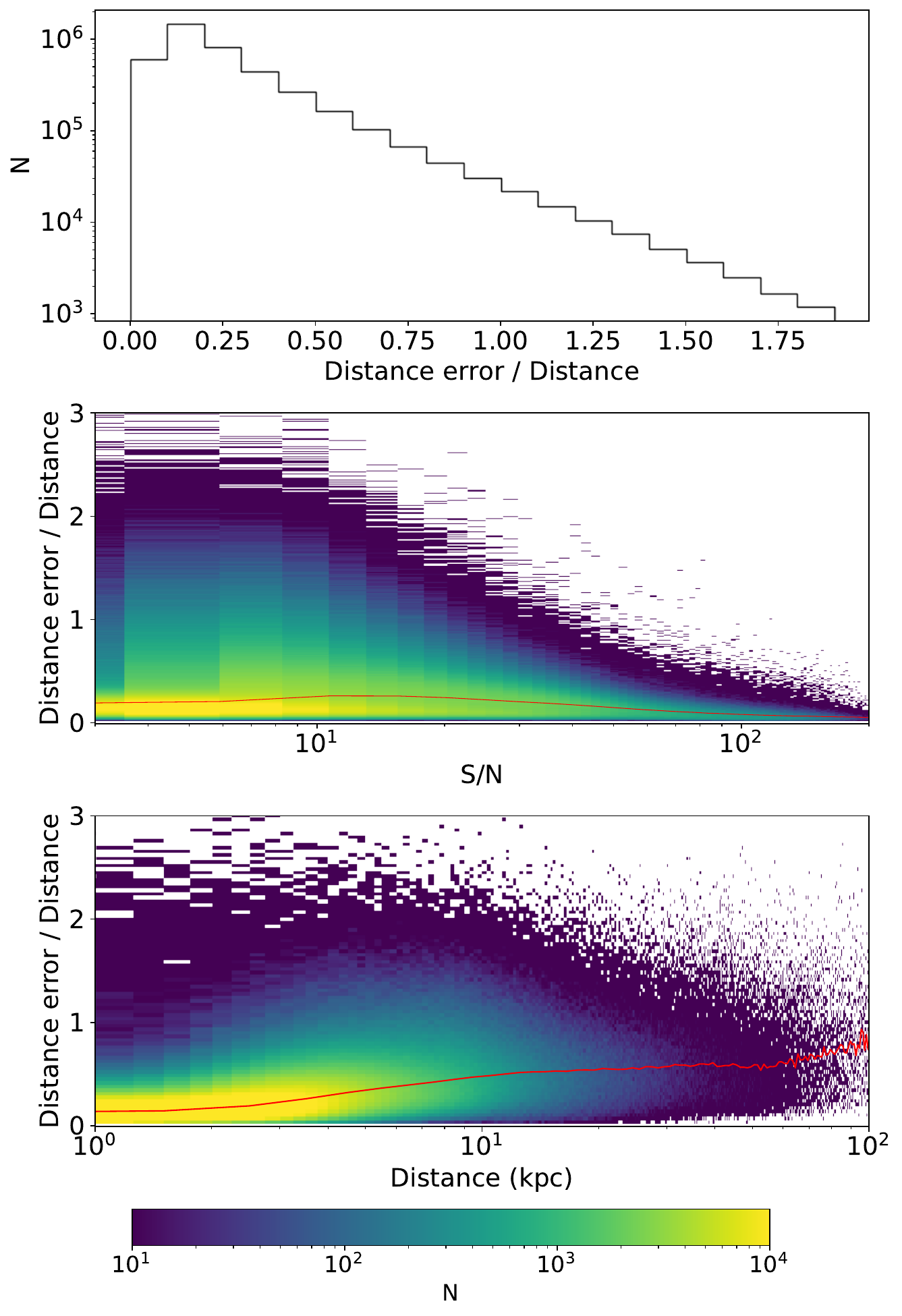}%
\end{center}
\caption{{\bf Top:} Distribution of relative distance uncertainty of the total sample. {\bf Middle}: relative distance uncertainty as a function of S/N. The red line shows the median value of the relative error at various S/N. {\bf Bottom:} Relative distance uncertainty of the total sample as a function of distance, the distance uncertainty is derived by the method of Section~\ref{sec:error model}. The red line shows the median value of the relative error at various distances. }
\label{fig:error distribution}
\end{figure}

Figure \ref{fig:error distribution} displays the distribution of relative distance uncertainty (top plot), relative distance uncertainty versus S/N of the total sample (middle plot) and the relation between relative distance uncertainty and distance (bottom plot). Here the uncertainties are all derived from the method\footnote{The uncertainties presented in Figure~\ref{fig:sn mag} are different. They are based on a subset of the test sample with precise {\it Gaia} parallax measurements ($\omega/\sigma_{\omega}>10$). It is also different from the external validation in Section~\ref{sec:Results on validation sets}.} in Section~\ref{sec:error model}. We find that the relative distance uncertainties of most stars are smaller than 50\%. About 60\% of the stars have a relative distance uncertainty of less than 25\%. 

The middle panel of Figure \ref{fig:error distribution} shows that the distance uncertainty is related to S/N. The relative distance uncertainty decreases with increasing S/N at S/N$>$10. The relative uncertainty is only 5\% when S/N is $\sim$200. These trends are consistent with Figure~\ref{fig:sn mag}. Specifically, the median distance uncertainties are 23\% for S/N $<$ 20, 19\% for 20 $\leq$ S/N $<$ 60, 11\% for 60 $\leq$ S/N $<$ 100, and 7\% for S/N $\geq$ 100.

The bottom panel of Figure~\ref{fig:error distribution} shows the relative distance uncertainty as a function of distance. The uncertainty is about 14\% within 2~kpc, which then quickly increases to the mean of $\sim$50\% at larger distances.

\subsection{Distance uncertainty for giant stars and comparison with SEGUE K giants}

The previous subsection (Section~\ref{sec:error estimation}) presents the distance uncertainty statistics for all stars. In this subsection we aim to quantify the uncertainties for giant stars. We start our analysis by performing a detailed comparison with SEGUE K-giants \citep{2014ApJ...784..170X}. 

The SEGUE distance catalog of 6,036 K-giants \citep{2014ApJ...784..170X} is not based on machine learning, and instead it is a probabilistic approach under the Bayesian framework. It relies on absolute magnitude versus color relations deduced from GCs with known distances and different metallicity. A median distance precision of 16\% is reported. The distance catalog can go up to 125~kpc from the Galactic center, with 283 stars beyond 50~kpc.

\begin{figure} 
\begin{center}
\includegraphics[width=0.4\textwidth]{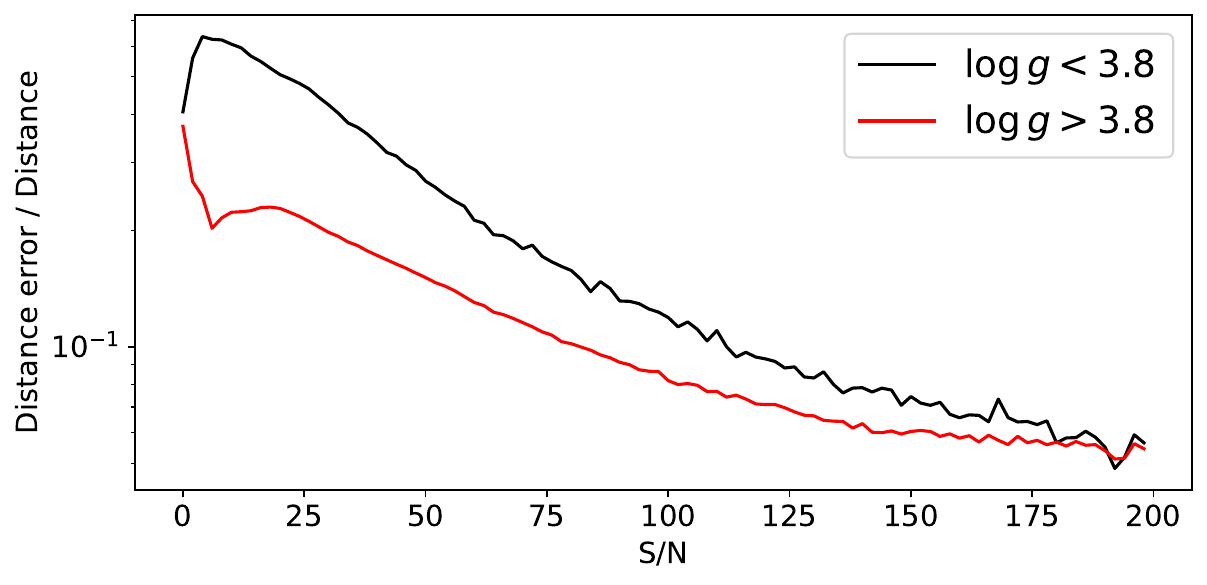}%
\end{center}
\caption{Median distance measurement errors obtained by Monte Carlo sampling the error spectra (see Section~\ref{sec:error model}), reported as a function of the S/N of the stellar spectra. Black and red curves are for stars with $\log g<3.8$ and $\log g>3.8$, respectively.}
\label{fig:errorforgiant}
\end{figure}

If we separate giant and main sequence stars with $\log g<3.8$ and $\log g>3.8$, our median distance uncertainty of giants can be as large as 40\% within 50~kpc, which is significantly larger than that of \cite{2014ApJ...784..170X}. 
Figure~\ref{fig:errorforgiant} shows the uncertainty versus S/N of stellar spectra, and we show this for stars with $\log g<3.8$ and $\log g>3.8$ separately. For giant stars with $\log g<3.8$, we can clearly see that the uncertainties are significantly larger than those of the other stars, indicating our model predicts much larger distance uncertainties for giant stars at fixed S/N. 
The median distance uncertainties for giant stars with $\log g<3.8$ are 58\% for S/N$<$20, 38\% for 20$\leq$S/N$<$ 60, 16\% for 60$\leq$ S/N $<$ 100 and 8\% for S/N$\geq$100.

Despite the large uncertainties for giant stars, if we restrict our sample to stars with $\log g<3.8$ and with distance uncertainties smaller than 25\%, we still have more than 74,000 and 1,500 stars within and beyond 50~kpc to the Galactic center, more than those in earlier surveys. This is benefited from the large DESI survey.

The top left panel of Figure~\ref{fig:comparison} shows a direct comparison between our measurements and those of \cite{2014ApJ...784..170X}. The number of matched stars is small, but the agreement is reasonable, in that the black dots go well through the red solid diagonal line, without biases. We further show in the top left panel of Figure~\ref{fig:comparison_ext} the external validation for \cite{2014ApJ...784..170X} based on our external validation sample (see Section~\ref{sec:validation set}). The validation looks very good, with small scatters and no particular biases. The amount of scatter based on this external validation is only 24\%.

The fact that our measurement uncertainty is larger than \cite{2014ApJ...784..170X} is related to three factors: 

1) DESI is much deeper, and thus even within 50~kpc, stars in DESI are fainter and have lower S/N in their stellar spectra. 

2) The method of \cite{2014ApJ...784..170X} is dedicated to K giants, whereas our model is trained to predict the distance for a variety of different types of stars. Although our distance measurements are unbiased up to 100~kpc, giants in our training sample contribute only 0.7\% in the loss function (see Section~\ref{sec:label_loss}) due to their smaller number and larger parallax errors. This can lead to a worse distance precision for giants than main sequence stars (see Figure~\ref{fig:errorforgiant}). The model that is trained to be optimized for the entire data set in our analysis may not perform equally well as a model or method particularly aimed for giant stars. We should train giants and main sequence stars separately in the future.

3) \cite{2014ApJ...784..170X} adopted member stars of GCs to calibrate the relation among absolute magnitudes, color and metallicity, whereas we only used member stars from GCs for our external validation, rather than training.

4) Lastly, stellar colors are not included in our training of this study. Continuum normalized stellar spectra have to be used to ensure proper convergence in the training process. However, this causes the loss of information about the stellar color and $T_\mathrm{eff}$ in the continuum. And most of the information in our training is from absorption lines.

Improvements associated with the items above will be incorporated in a future study with DESI Year-3 data.

\subsection{Comparison with other recent measurements}

\begin{figure*} 
\begin{center}
\includegraphics[width=0.8\textwidth]{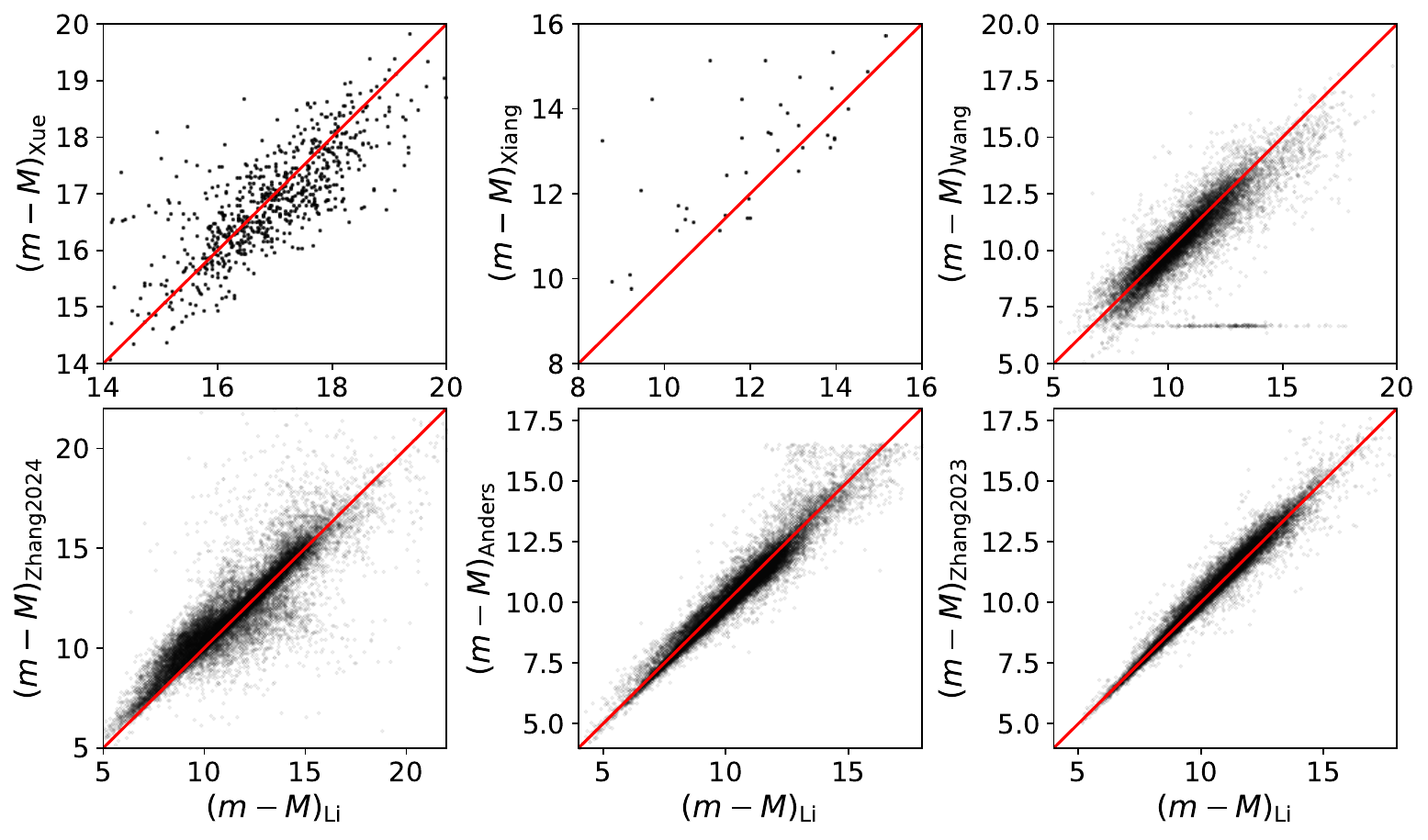}%
\end{center}
\caption{Comparison between our distance measurements in this paper ($x$-axis) and the spectrophotometric distance measurements in six previous studies, including SEGUE K-giants \citep[][top left]{2014ApJ...784..170X}, LAMOST OB stars \citep[][top middle]{2021ApJS..253...22X}, LAMOST value-added catalog \citep[][top right]{2022ApJS..259...51W}, the more recent measurement for DESI EDR based on the Payne \citep[][bottom left]{2024ApJS..273...19Z}, photo-astrometric distances of {\it Gaia} EDR3 \citep{starhorse2} (bottom middle), and distances measured from {\it Gaia} BP/RP spectra \citep[][bottom right]{GaiaXp}. For top right, bottom middle and bottom right panels, we only compare those stars which are flagged as good measurements. In all panels, the red solid line marks $y=x$ to guide the eye.}
\label{fig:comparison}
\end{figure*}

\begin{figure*} 
\begin{center}
\includegraphics[width=0.8\textwidth]{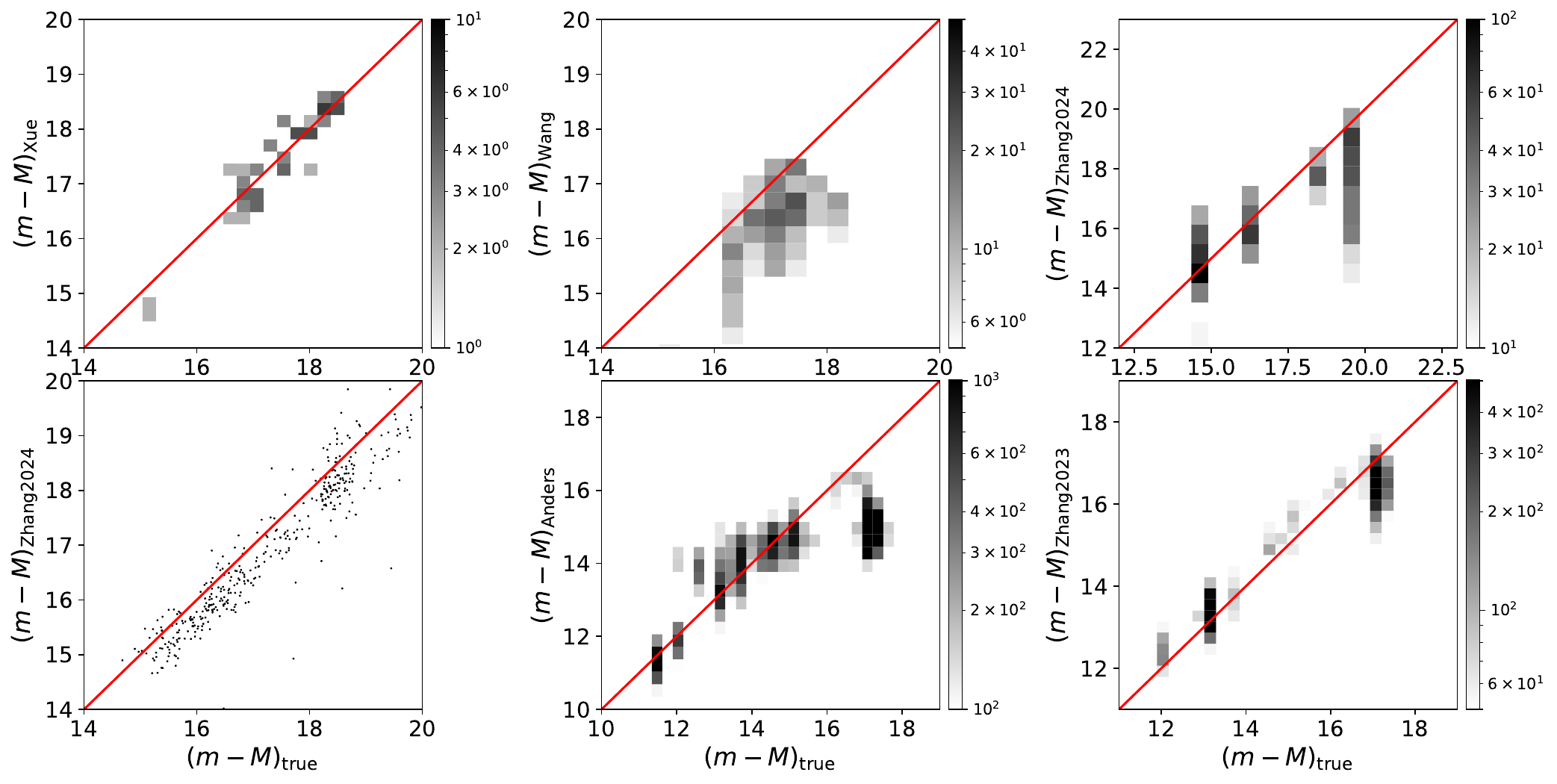}%
\end{center}
\caption{External validations for the distance measurements of \cite{2014ApJ...784..170X} (top left), \cite{2022ApJS..259...51W} (top middle), \cite{2024ApJS..273...19Z} (top right and bottom left), \cite{starhorse2} (bottom middle) and \cite{GaiaXp} (bottom right). For the top left and middle panels, most stars are matched to the member star list of Sgr. The top right panel is based on matched member stars to a few GCs and dwarf galaxies. The bottom left panel is based on DESI Year-1 BHB stars. For the bottom middle and right panels, we only present the validation based on GCs and Sgr member stars. For top middle, bottom middle and right panels, only stars flagged as good measurements are plotted. In all panels, the red solid line marks $y=x$ to guide the eye.}
\label{fig:comparison_ext}
\end{figure*}

In this subsection, we perform comparisons with a few other most recent and relevant spectrophotometric distance measurements, including LAMOST OB stars \citep{2021ApJS..253...22X}, the LAMOST value-added stellar parameter catalog \citep{2022ApJS..259...51W}, the more recent measurements for DESI EDR data based on the Payne \citep{2024ApJS..273...19Z}, the {\it Gaia} EDR3 stellar catalog \citep{starhorse2} base on StarHorse and the {\it Gaia} DR3 stellar catalog measured from XP spectra \citep{GaiaXp}.

The methodology of our current paper largely follows\footnote{Note \cite{2021ApJS..253...22X} also adopts PCA in their analysis. PCA has been used to remove emission lines from their OB stars.} \cite{2021ApJS..253...22X}. 
The distance measurements by \cite{2021ApJS..253...22X} focus on nearby OB stars within 15~kpc to the Sun and from LAMOST. The precision in the measured distances of \cite{2021ApJS..253...22X} is $\sim$12\%. At S/N$>$20 and within 15~kpc, our precision is 14\%, which is comparable to \cite{2021ApJS..253...22X}, but we do not limit to only OB stars.

The measurements of \cite{2022ApJS..259...51W}, \cite{2024ApJS..273...19Z}, \cite{starhorse2} and \cite{GaiaXp} all cover wide stellar types, similar to ours. All of them provide measurements of stellar parameters, with the distance measurement a byproduct.

\cite{2022ApJS..259...51W} claimed 8.5\% precision for S/N $>$ 50, while our precision is 10\% at S/N $>$ 50, so comparable. The measurements of \cite{2024ApJS..273...19Z} were made using the data-driven Payne \citep{2019ApJ...879...69T,2019ApJS..245...34X,2022A&A...662A..66X}. Their precision is 14\% for 5 $<$ S/N $<$ 10 and smaller than 4\% for S/N >50. Our precision is 19\% for 5$<$ S/N $<$10 and 10\% for S/N $>$ 50. 
\cite{starhorse2} measured the stellar parameters for {\it Gaia} EDR3 using StarHorse. The distance precision of \cite{starhorse2} is 2\% for {\it Gaia} $G$-band apparent magnitude brighter than 14, 8\% around $G\approx16$ and 20\% around $G\approx20$. Our precision is 7\% for $G<14$, 15\% around $G\approx16$, 48\% around $G\approx20$.

Similar to the Payne, \cite{GaiaXp} builds forward models and fits the stellar parameters to the {\it Gaia} XP spectra to estimate stellar parameters and distances. 
They claimed that the typical uncertainty of predicted parallax is 0.01 mas. Transforming this typical uncertainty of parallax to distance precision, the distance precision is only 1\% at $\sim$1~kpc, 15\% at $\sim$15~kpc. The distance precision of our VAC is 10\% at $\sim$1~kpc, and 34\% at $\sim$15~kpc.

Overall speaking, \cite{2024ApJS..273...19Z},\cite{starhorse2} and \cite{GaiaXp} report smaller distance uncertainties than this VAC, and later in this subsection, we will show external validation results of \citep{2024ApJS..273...19Z,starhorse2,GaiaXp}.

The top middle, top right and three bottom panels of Figure~\ref{fig:comparison} show direct comparisons between our distance measurements and those of \cite{2021ApJS..253...22X}, \cite{2022ApJS..259...51W}, \cite{2024ApJS..273...19Z}, \cite{starhorse2} and \cite{GaiaXp}, based on cross matched stars between our VAC and the public catalogs of these previous studies. For those catalogs \citep{2022ApJS..259...51W, starhorse2, GaiaXp} which provide distance quality flags, we only plot the stars flagged as good.

In general, our measurements show reasonable agreements with all the five previous studies. There are, however, some detailed differences. In the top middle panel, where we show the comparison with \cite{2021ApJS..253...22X}, the black dots tend to be more above the red diagonal line. However, this is perhaps associated with the small sample of stars matched, so we avoid having strong comments on this. 

In the top right and bottom left panels where the samples of matched stars are much larger, the measured distance modulus by \cite{2022ApJS..259...51W} tends to fall below the diagonal line than ours beyond 14. The measurements by \cite{2024ApJS..273...19Z} go above the diagonal line than ours\footnote{Here we have used a subset of stars with precise {\it Gaia} parallax ($\omega/\sigma_\omega>10$) to validate, and find our measurements are unbiased over $5<(m-M)_\mathrm{Li}<12$.} over distance modulus range of $5<(m-M)_\mathrm{Li}<12$. 

For the bottom middle and bottom right panels, our distance measurements agree with \cite{starhorse2,GaiaXp} quite well over the distance modulus range of $5<(m-M)_\mathrm{Li}<15$. The comparison is not clear 
at $(m-M)_\mathrm{Li}>15$ due to the small number of matched stars.

We further conduct external validations of these previous studies by matching the sample of \cite{2022ApJS..259...51W, 2024ApJS..273...19Z, starhorse2,GaiaXp} with the external validation set in Section~\ref{sec:validation set}. For \cite{2022ApJS..259...51W}, \cite{starhorse2} and \cite{GaiaXp}, we only present the external validations for stars which have good flags. We do not show the validation for \cite{2021ApJS..253...22X}, as there are few matched stars to the external validation sample.

The top middle panel of Figure~\ref{fig:comparison_ext} shows the validation of \cite{2022ApJS..259...51W}. The distance measurements by \cite{2022ApJS..259...51W} are subject to some underestimates, with more data below the diagonal line. The top right and bottom left panels of Figure~\ref{fig:comparison_ext} show the external validations for the distance measurements of \cite{2024ApJS..273...19Z}. The top right panel is based on member stars of a few GCs, Sgr member stars and dwarf galaxies. There is no particular bias for distance modulus of $15<(m-M)_\mathrm{true}<18$, but at $(m-M)_\mathrm{true}$ close to 20 (100~kpc), there are some underestimates. The bottom left panel of Figure~\ref{fig:comparison_ext} shows the validation using DESI Year-1 BHB stars. It seems there is some slight but almost constant underestimate. 

The bottom middle and right panels of Figure~\ref{fig:comparison_ext} show the external validations for the distance measurements of \cite{starhorse2} and \cite{GaiaXp}. Here we only present the validations based on matched GCs and Sgr member stars. The matched number of member stars in other dwarf galaxies is very few. Both panels show approximately unbiased distances up to $(m-M)_\mathrm{true}\sim16$, but the distances are more underestimated at $(m-M)_\mathrm{true}>16.5$ ($>$20~kpc).

\section{Binary identification}
\label{sec:Binary identify}

Unresolved binaries can be identified through radial velocity variations \citep[e.g.][]{1996ASPC...90...21M,2011AJ....141..200M,2017ApJ...837...20P,2018RAA....18...52T,2018AJ....156...18P,2020ApJ...895....2P,2020ApJ...905...67P} or spectral double lines \citep[e.g.][]{2010AJ....140..184M,2017A&A...608A..95M,2021AJ....162..184K,bianry_lamost_22}, but these methods require multiple observations or high-resolution spectra. Alternatively, binaries can be identified through the HR diagram\citep[e.g.][]{2010MNRAS.401..577S,2011AJ....141..115C,2012A&A...540A..16M,lamost_binary, Li_lu_20}, because binary sequences appear above the main sequence due to brighter luminosity. By fitting the difference between $M_{G,\mathrm{geo}}$ and the magnitude along the ridge line for single stars ($M_{G,\mathrm{ridge}}$), binaries can be distinguished from single stars.
In this work, we can also identify binaries by fitting the difference between $M_{G,\mathrm{geo}}$ and $M_{G,\mathrm{NN}}$. This approach retains the same key idea as using the HR diagram.

However, the currently proposed model is effective only for the stars with small measurement errors in their parallaxes, like a few previous works \citep[e.g.][]{lamost_binary, Li_lu_20}. Moreover, we expect this method to work well with main sequence binary systems of nearly equal-mass, but not for giants, whose companions are mostly dwarfs. We first introduce and discuss our proposed modeling approach below, and present some initial results at this stage. 

\begin{figure} 
\begin{center}
\includegraphics[width=0.4\textwidth]{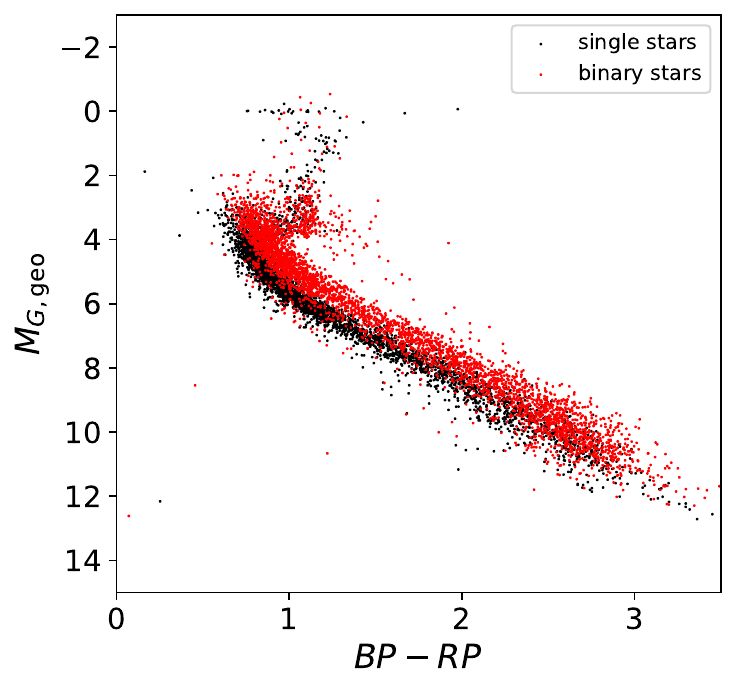}%
\end{center}
\caption{The distribution of single stars (black dots) and binaries (red dots) in Hertzsprung-Russell diagram, and binaries are identified by our Gaussian mixture model (Equation~\ref{equ:binary possibility}) with $P_\mathrm{binary}>0.7$. Stars plotted in the figure are based on a subset of test sample with precise parallax measurements of $\omega/\sigma_{\omega}>10$. The $x$-axis is the color index based on {\it Gaia} photometry in $BP$ and $RP$ bands. The $y$-axis is the geometric absolute magnitude $M_{G,\mathrm{geo}}$ based on {\it Gaia} parallax. The distribution of binaries shows a clear shift from single stars by $\sim$0.75 magnitude.}
\label{fig:hr binary single}
\end{figure}

As demonstrated in Figure~\ref{fig:sn mag}, equal-mass binaries show an offset of $\sim$0.75 magnitude above the diagonal line. The relation between stellar luminosity and mass obeys a power law \citep{l-m-relation}. If assuming the primary and secondary stars in a binary system obey a cubic power-law relationship between luminosity and mass, the difference between $M_{G,\mathrm{geo},\mathrm{binary}}$ and $M_{G,\mathrm{geo},\mathrm{single}}$, which is also the difference between $M_{G,\mathrm{geo},\mathrm{binary}}$ and $M_{G,\mathrm{NN}}$, can be written as:
\begin{equation}
\label{equ:mass-ratio-to-delta_mag}
M_{G,\mathrm{geo},\mathrm{single}} - M_{G,\mathrm{geo},\mathrm{binary}} = 2.5(\log{(1 + \rho^3)}),
\end{equation}
where $\rho$ denotes the mass ratio of the secondary star to the primary star.

Typically, $M_{G,\mathrm{geo},\mathrm{binary}}$ is 0.45 to 0.75 magnitudes brighter than $M_{G,\mathrm{NN}}$ for binaries with a mass ratio ranging from 0.8 to 1. The differences between $M_{G,\mathrm{NN}}$ and $M_{G,\mathrm{geo}}$ of binaries are large enough to be detected by the NN, as the typical error of the NN is much smaller than 0.45. Therefore, we can construct a new model to detect binaries.

A stellar spectrum may originate from a single star or a binary system, and thus the corresponding absolute magnitude should not be considered as a single value but rather as a distribution. We assume that the absolute magnitude obeys a Gaussian mixture distribution with two Gaussian components, with one component representing single stars and another component representing binaries. Notably, in our modeling here for binaries, we utilize absolute magnitudes as our training label. 

\begin{figure} 
\begin{center}
\includegraphics[width=0.24\textwidth]{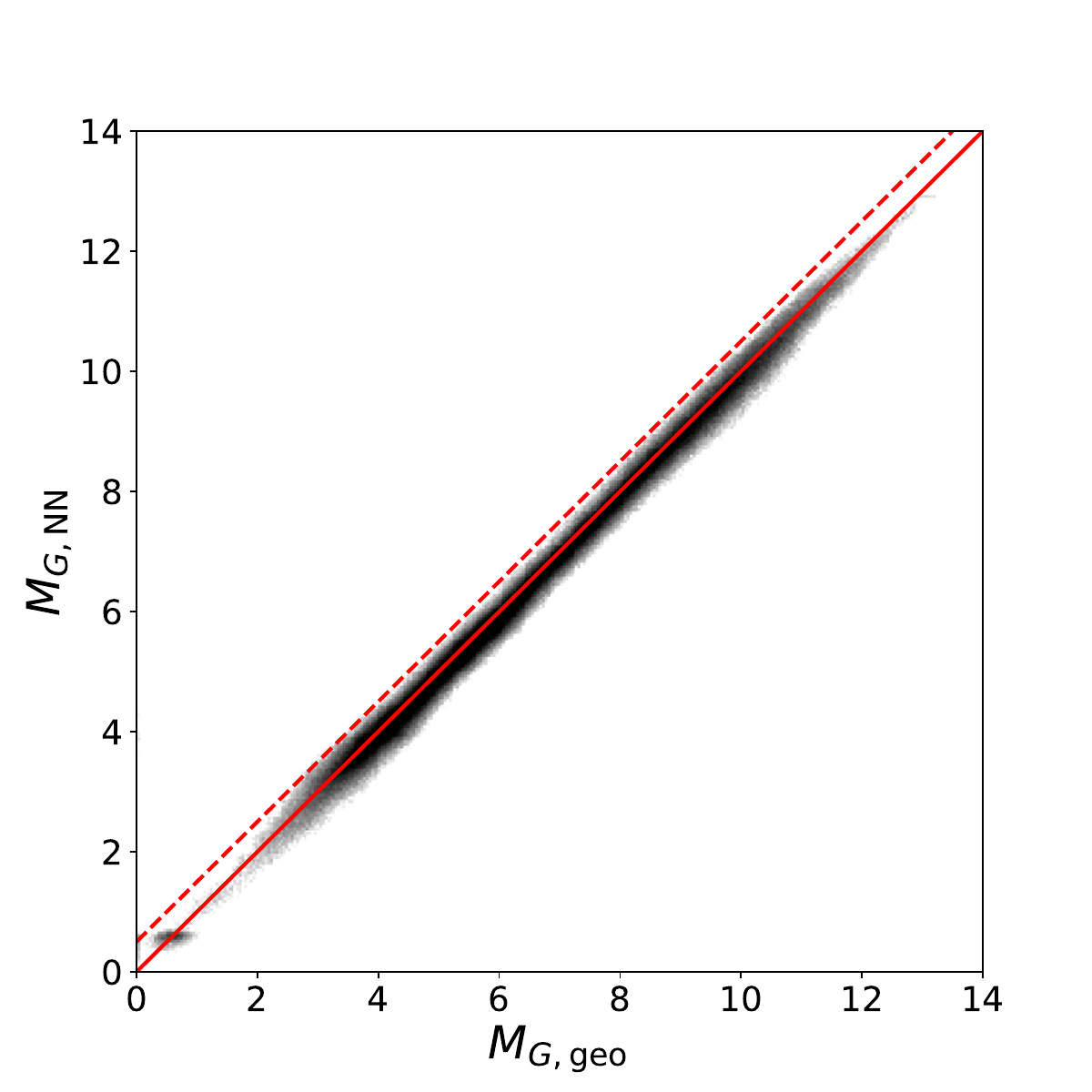}%
\includegraphics[width=0.24\textwidth]{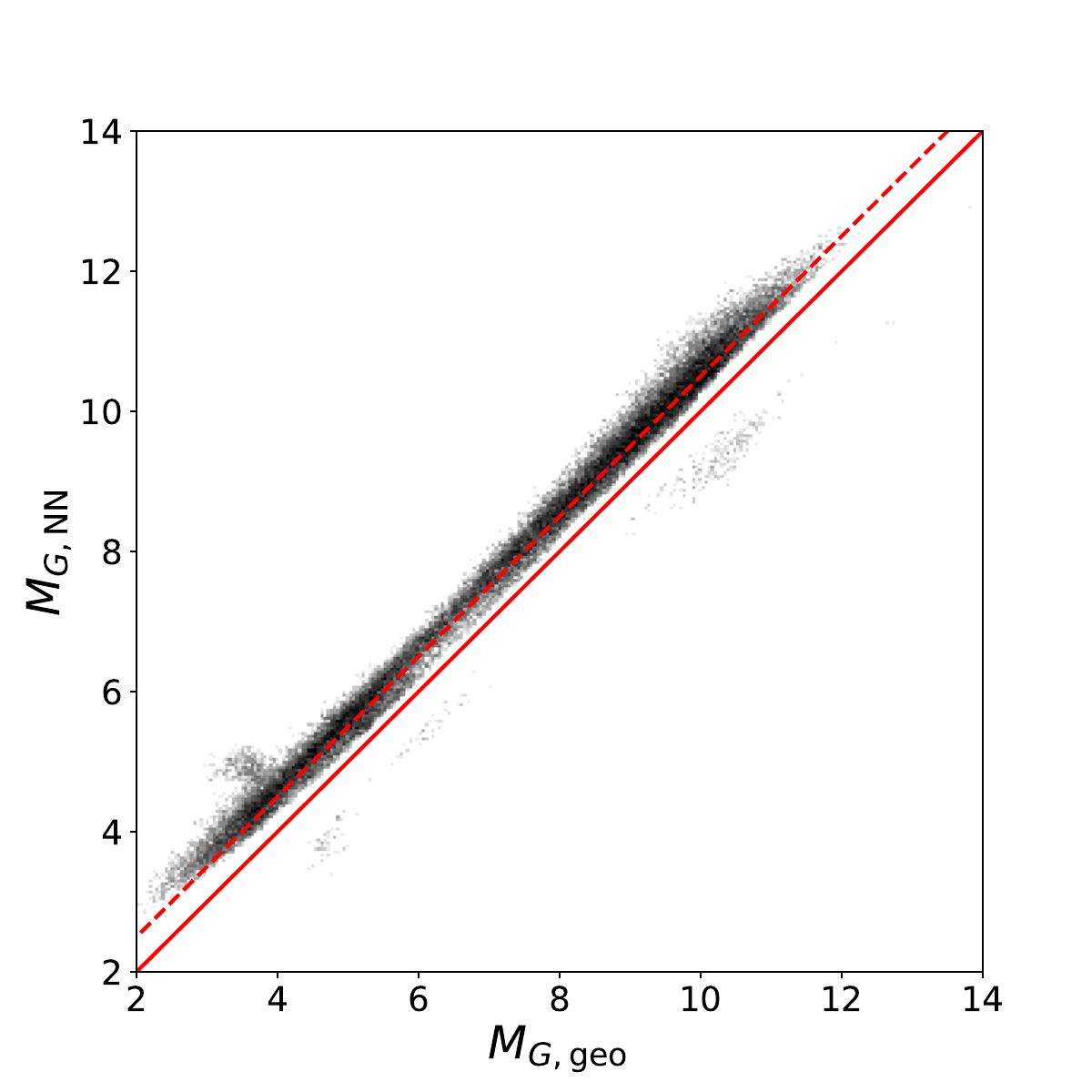}%
\end{center}
\caption{Comparison between $M_{G,\mathrm{NN}}$ and $M_{G,\mathrm{geo}}$. Here $M_{G,\mathrm{NN}}$ is obtained in Section~\ref{sec:Results of training and test sets}, and we only plot a subset of the test sample with precise parallax measurement $\omega/\sigma_{\omega}>10$, based on which that we can deduce precise $M_{G,\mathrm{geo}}$. We apply the method of Section~\ref{sec:Binary identify} to distinguish binaries and single stars. The left plot is for single stars, and the right plot is for binary stars. In both panels, the red solid line shows the $y=x$ diagonal line, and the red dashed line has an offset of 0.75 magnitude above the diagonal line.}
\label{fig:binary identification}
\end{figure}

For a given stellar observation, the Gaussian mixture model describing the distribution of absolute magnitude, $M$, has the following form:

\begin{equation}
\label{equ:two gaussian mixture}
P(M|\Theta) = \alpha N_1(M-\mu_1|\sigma_1) + (1 - \alpha) N_2(M-\mu_2|\sigma_2),
\end{equation}
and we utilize NN to predict the five model parameters (denoted as $\Theta=\{\alpha, \mu_1, \sigma_1, \mu_2, \sigma_2\}$) of the two Gaussian distributions ($N_1$ and $N_2$) for each stellar observation with standard deviations ($\sigma_1$ and $\sigma_2$), mean values ($\mu_1$ and $\mu_2$) and the fraction of each component ($\alpha$). In Equation~\ref{equ:two gaussian mixture}, the first and second Gaussian components refer to single and binary stars, respectively. Here we enforce $\mu_1$ to be greater than $\mu_2$ as the binary component has a brighter luminosity which leads to a smaller $\mu$. 

The likelihood can be constructed by convolving the Gaussian mixture distribution with the observational uncertainty of absolute magnitudes:

\begin{equation}
\label{equ:likelihood}
\mathrm{Likelihood} = \prod_i \int P(M|\Theta_i)P(M|\omega_{0,i},m_{G,i}) dM,
\end{equation}
where $P(M|\Theta_i)$ is the Gaussian mixture distribution for star $i$, i.e., Equation~\ref{equ:two gaussian mixture}. The second term of $P(M|\omega_{0,i},m_{G,i})$ is the error model for the absolute magnitudes. Instead of assuming the error of absolute magnitude is Gaussian, we assume that the directly measured parallax error is Gaussian, and we deduce the error model for absolute magnitude from the Gaussian error of parallax as $P(M|\omega_{0,i},m_{G,i})=N(10^{\left(\frac{M-m_{G,i}+10}{5}\right)}-\omega_{0,i}|\sigma_{\omega_{0,i}})10^{\left(\frac{M-m_{G,i}+10}{5}\right)}$. Here $m_{G,i}$, $\omega_{0,i}$ and $\sigma_{\omega_{0,i}}$ are the observed {\it Gaia} $G$-band apparent magnitude, parallax and parallax error for the $i$-th star, which are fixed values. $i$ in Equation~\ref{equ:likelihood} goes for each individual star, and the parameters of the Gaussian mixture model differ for each individual star. 

After obtaining the best trained NN, we can estimate the probability for a given star of being a binary system as:

\begin{equation}
\label{equ:binary possibility}
P_\mathrm{binary,i} = \frac{\int (1 - \alpha_i) N_2(M-\mu_{2,i}|\sigma_{2,i})P(M|\omega_{0,i},m_{G,i})dM}{\int P(M|\Theta_i)P(M|\omega_{0,i},m_{G,i})dM}.
\end{equation}

We apply the aforementioned modeling approach to a subset of stars with precise parallax measurements ($\omega/\sigma_{\omega}>10$), and use Equation~\ref{equ:binary possibility} to estimate the probability for each star to be a binary system ($P_\mathrm{binary,i}$). Due to the relative error cuts applied to the parallax, most of the selected stars in the subsample are nearby main sequence stars located within 5~kpc. We classify a star as a binary if $P_\mathrm{binary,i}$ is greater than 0.7. In this test sample, about 21,000 stars satisfy this criterion and are considered to be binaries. Figure~\ref{fig:hr binary single} shows the distribution of these binaries identified in this way in red, and black dots are single stars. The distribution of binaries moves upwards by $\sim$0.75 magnitude, indicating they are mostly equal-mass binaries.

Figure~\ref{fig:binary identification} further compares $M_{G,\mathrm{geo}}$ and $M_{G,\mathrm{NN}}$, and here $M_{G,\mathrm{NN}}$ is the NN predicted absolute magnitudes that we obtained in Section~\ref{sec:results} (not based on the binary model in the current section). The figure is also based on the subset of stars with precise parallax measurements. The left and right plots are for stars with $P_\mathrm{binary}\leq0.7$ and $P_\mathrm{binary}>0.7$. As can be clearly seen, we are able to successfully distinguish the single and binary regions with our modeling approach, when we have precise parallax measurements. 

In particular, the left panel of Figure~\ref{fig:binary identification} shows that the precisions in $M_{G,\mathrm{NN}}$ for single stars are 0.19 and 0.26 magnitudes for S/N $>$ 20 and S/N $<$ 20, which corresponds to precisions of 9\% and 13\% in distance respectively. Note if not excluding binaries, the precisions in distance measurements are 10\% and 18\% for S/N $>$ 20 and S/N $<$ 20. 

\begin{figure}
\begin{center}
\includegraphics[width=0.4\textwidth]{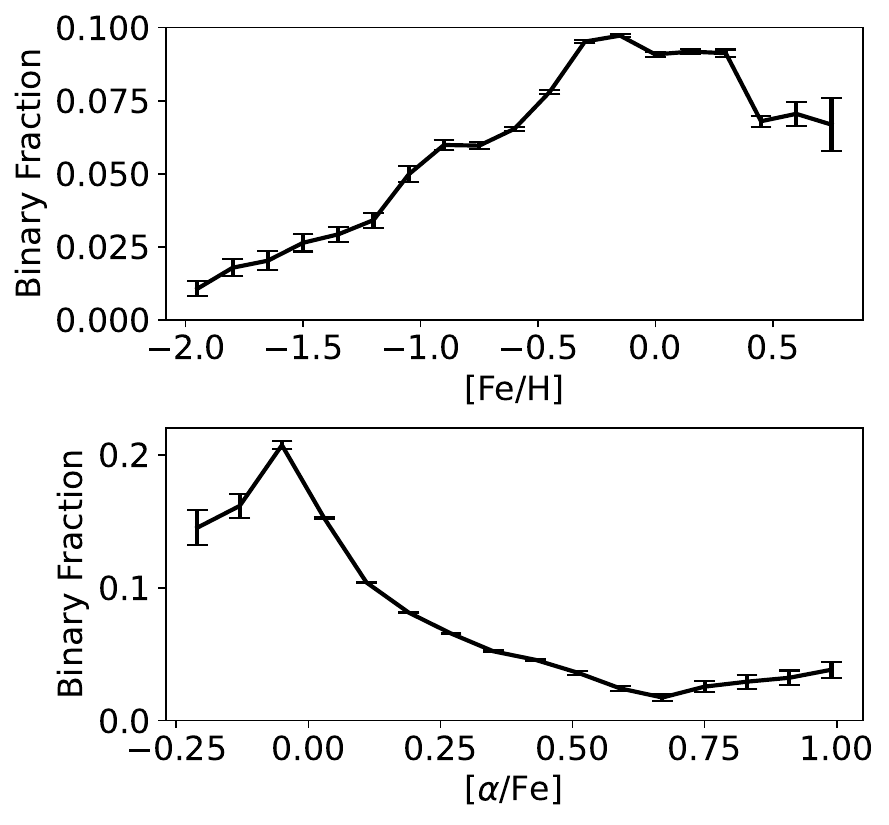}%
\end{center}
\caption{Equal-mass binary fraction reported as a function of [Fe/H] and [$\alpha$/Fe]. This is based on a subset of stars having precise {\it Gaia} parallax measurements, i.e., $\omega/\sigma_{\omega}>10$. Errorbars represent the 1-$\sigma$ scatter among 50 bootstrap subsamples. }
\label{fig:binary fraction}
\end{figure}

In the right panel of Figure~\ref{fig:binary identification}, $\sim$10\% of the identified binary stars are located below the diagonal line by $\sim$0.5 magnitudes. These stars are distinguished and modeled by the second component of our mixture model. However, they are unlikely to be true binary systems and may instead be contaminants, such as white dwarfs. Indeed, about half of these stars do have the colors of white dwarfs. The S/N is very low for the other half, and the reason why they locate below the diagonal line may be due to measurement uncertainties. In order to remove those misclassified binaries, we further exclude those with $M_{G,\mathrm{NN}}$ brighter than $M_{G,\mathrm{geo}}$. Our final binary criteria is:
\begin{equation}
\label{equ:final criteria}
\left\{\begin{aligned}
P_\mathrm{binary}&>0.7,\\
M_{G,\mathrm{NN}}&>M_{G,\mathrm{geo}},
\end{aligned}\right.
\end{equation}

We have also applied this modeling approach above to the full sample of stars without cuts in parallax error. However, the final converged estimates of the model parameters become unphysical, with very large $\sigma_{1,2}$ at brighter magnitudes and the recovered $\mu_{1,2}$ being more different from the measurements in Section~\ref{sec:results}. 
We attribute this primarily to the fact that giants dominate brighter magnitudes, and their companions are predominantly dwarfs, leading to smaller mass ratios. Moreover, the parallax measurements of these more distant stars suffer from very large uncertainties at brighter luminosities (fainter apparent magnitudes as they are more distant). The large uncertainties not only hinder the ability of the model to discern a clear double Gaussian distribution, but also suggest that the error distribution of the observed {\it Gaia} parallaxes may not be Gaussian. We acknowledge this limitation and leave more detailed investigations about how to more accurately identify binary stars and improve our distance measurements with the full sample to future studies.

With the full sample, 120,000 stars are identified as binaries with Equation~\ref{equ:final criteria}. Among the 120,000 binary candidates, 90,000 have precise {\it Gaia} parallax measurements, satisfying $\omega/\sigma_{\omega}>10$. 
We have checked the distribution of all these 120,000 binary star candidates, and find that most of their $M_{G,\mathrm{NN}}$ obtained in Section~\ref{sec:results} based on the single star model is indeed about 0.5-0.75 magnitudes fainter than $M_{G,\mathrm{geo}}$, and thus the identification of these 120,000 stars as binaries is robust, with most of them equal-mass binary systems. In our final \textsc{SpecDis} VAC product (see Table~\ref{tab:table}), we include a column to indicate these 120,000 binary candidates. 

In our VAC, we also provide a column that gives the binary possibility of stars in the full sample. However, we warn the readers that even when $P_\mathrm{binary}$ is close to unity, there can be some contamination. This happens mainly when the uncertainty of $M_{G,\mathrm{NN}}$ (based on the single star model) or $M_{G,\mathrm{geo}}$ is large. If the readers also adopt Equation~\ref{equ:final criteria}, it is fine, as the contamination can be largely eliminated by $M_{G,\mathrm{NN}}>M_{G,\mathrm{geo}}$. However, if the readers adopt other criterion in $P_\mathrm{binary}$ more different from 0.7, we recommend additionally including the following selection:
\begin{equation}
\label{equ:binary choose criteria}
M_{G,\mathrm{NN}} - M_{G,\mathrm{geo}} > \sqrt{M_{G,\mathrm{NN},\mathrm{error}}^2 + M_{G,\mathrm{geo},\mathrm{error}}^2}.
\end{equation}

Based on the identified 120,000 equal-mass binaries, Figure~\ref{fig:binary fraction} shows the dependence of the binary fraction on [Fe/H] and [$\alpha$/Fe]. Here we only present results for the subsample that has precise parallax measurements of $\omega/\sigma_{\omega}>10$. We calculate the equal-mass binary fraction as the ratio between the identified binaries versus the total number of stars in this subsample.

Figure~\ref{fig:binary fraction} shows that the equal-mass binary fraction increases with [Fe/H] and declines with [$\alpha$/Fe]. [Fe/H] and [$\alpha$/Fe] are anti-correlated, so the correlation between the equal-mass binary fraction and [Fe/H] or [$\alpha$/Fe] is likely due to the same reason, that for equal-mass binaries having lower [Fe/H] (more metal poor) or higher [$\alpha$/Fe] (more rapid star formation\footnote{Core-collapse supernovae are triggered soon after the onset of star formation, release mostly $\alpha$ elements. [$\alpha$/Fe] thus reflects the star formation time scale \citep[e.g.][]{1987A&A...185...51M,1999MNRAS.302..537T,2003MNRAS.344..455F,2005ApJ...621..673T,2010MNRAS.404.1775T,2020MNRAS.497.3557L}.}), they form earlier in the Universe. Stars formed earlier have a longer time to experience possible perturbations such as close encounters with another object, resulting in disruptions of the equal-mass binary systems, hence the equal-mass binary fraction becomes lower.

However, our results show an opposite trend from a few previous studies. For example, \cite{2010ApJS..190....1R} claimed that metal-poor systems have a higher binary fraction than metal-rich systems. \cite{2018RAA....18...52T} claimed that $\alpha$ abundances are positively correlated with binary fractions. Moreover, \cite{lamost_binary} assumed that the mass-ratio distribution of binaries obeys a power law with a power index of $\gamma$, and found that the binary fraction of the high-$\gamma$ binary systems is anti-correlated with [Fe/H] while the binary fraction of the low-$\gamma$ binary systems is quite flat with [Fe/H]. 
The different trend shown in this study may be related to two factors: 1) \cite{2010ApJS..190....1R} and \cite{lamost_binary} focus on stars within 25~pc and 333~pc, respectively, whereas the median distance of binaries identified in our current study is $\sim$520~pc. Some of our identified binaries could even extend to $\sim$3~kpc. Binary fractions can vary in different environments, and with a larger distance range, our binary sample could be mixed with GSE, thin and thick disk components, leading to a different trend compared with \cite{2010ApJS..190....1R} and \cite{lamost_binary}.
2) We mainly identify equal-mass binaries, so the trend may not be generalized to binaries with mass ratios much smaller than unity. The trend could be different after taking non-equal-mass binaries into account. We leave more detailed studies on the binary fraction of non-equal-mass binaries to future studies.

\section{Conclusions}
\label{sec:concl}

In this paper, we have successfully implemented a data-driven method to measure the spectrophotometric distances for stars from the DESI Year-1 data (\textsc{SpecDis}). Moreover, we have estimated distance uncertainties by Monte-Carlo sampling the stellar spectra according to the error spectra. \textsc{SpecDis} involves training an NN on a large sample of stars without applying selection on either {\it Gaia} parallax error or S/N of the stellar spectra, ensuring a broad and unbiased training sample. We have employed PCA to reduce the noise and dimensionality of the input stellar spectra. Our NN predicts a label constructed from the reciprocal of the square root of luminosity, which is linearly proportional to parallax, thus ensuring that the prediction is unbiased even with a large parallax error.
Moreover, by incorporating the error in {\it Gaia} parallax into our loss function, we allow the NN to account for the uncertainties in parallax.

The precision in our measured distance improves with S/N for the stellar spectra. Based on Monte Carlo sampling the errors of the input spectra (see Section~\ref{sec:error model}), we find that the median distance uncertainties for different S/N ranges are as follows: 23\% for S/N $<$ 20, 19\% for 20 $\leq$ S/N $<$ 60, 11\% for 60 $\leq$ S/N $<$ 100, and 7\% for S/N $\geq$ 100. Giant stars with $\log g<3.8$ are subject to much larger uncertainties of 58\% for S/N$<$20, 38\% for 20$\leq$S/N$<$ 60, 16\% for 60$\leq$ S/N $<$ 100 and 8\% for S/N$\geq$100. After selecting giant stars with $\log g<3.8$ and with distance uncertainties smaller than 25\%, we still have more than 74,000 and 1,500 stars within and beyond 50~kpc to the Galactic center, benefiting from the large DESI MWS.

External validations are performed by using member stars from GCs, stellar streams and dwarf galaxies, which have precise distances and can be utilized as independent references. We also adopted the DESI Year-1 BHB stars for the validation. The external validation shows that there is no particular bias in our distance measurements within 100~kpc after applying an appropriate error cut.

Detailed comparisons are performed between our distance measurements and a few most relevant studies of \cite{2014ApJ...784..170X}, \cite{2021ApJS..253...22X}, \cite{2022ApJS..259...51W}, \cite{2024ApJS..273...19Z}, \cite{starhorse2} and \cite{GaiaXp}. In general, good agreements are found with these previous measurements. 

Our final \textsc{SpecDis} VAC product provides distance and distance uncertainty measurements for over 4 million stars, offering a valuable resource for Galactic astronomy and near-field cosmology studies. The catalog covers a wide range of distances, with 96~\% stars located within 15~kpc from the Sun, and about 120,000 stars located beyond 15~kpc. Some stars can even extend to 250~kpc.

We have also developed a method for identifying equal-mass binary systems by modeling the absolute magnitude distribution as a mixture of two Gaussian components. This approach has shown promise for a subset of stars with precise parallax measurements, allowing us to potentially improve the precision of our distance measurements and has identified 120,000 possible equal-mass binaries. However, this method is currently limited to main sequence stars with precise parallaxes. 

With the identified binaries, we found that the equal-mass binary fraction increases with [Fe/H] and declines with [$\alpha$/Fe]. This likely implies that equal-mass binaries that form earlier, which are hence poorer in Fe and richer in $\alpha$, they may have experienced more close encounters and tidal effects, hence the binary systems are disrupted more.

In summary, our work presents a robust and comprehensive distance VAC product for a large sample of stars from the DESI Year-1 data, demonstrating the power of data-driven approaches in Galactic astronomy. The catalog is released as one of the DESI DR1 VAC products at \url{https://data.desi.lbl.gov/doc/releases/dr1/vac/mws-specdis/}. The content and columns of our final \textsc{SpecDis} VAC product are provided in Table~\ref{tab:table}. There is another stellar distances VAC (\textsc{SPDist}) of DESI DR1, which can be found at \url{https://data.desi.lbl.gov/doc/releases/dr1/vac/mws-spdist/}. There will be a paper (Guillaume, et al., in prep) presenting comparisons between \textsc{SpecDis} and \textsc{SPDist}. 

The measurements presented in this paper can be accessed at \url{https://zenodo.org/records/14903565}, which contains all data points for figures presented in this work.

\begin{table*}
\centering
\scriptsize 
\begin{tabular}{|l|l|}
\hline
Field & Description \\
\hline
TARGETID & DESI source ID \\
SOURCE\_ID & Gaia DR3 source ID \\
RA & Gaia DR3 Right Ascension (deg) \\
DEC & Gaia DR3 Declination (deg) \\
PMRA & Gaia DR3 Proper Motion in Right Ascension \\
PMRA\_ERR & Uncertainty in pmra \\
PMDEC & Gaia DR3 Proper Motion in Declination \\
PMDEC\_ERR & Uncertainty in pmdec \\
VRAD & Radial velocity (km/s) \\
DIST & Heliocentric distances derived from $M_{G,\mathrm{NN}}$ (kpc) \\
DISTERR & Uncertainty of distance (kpc) \\
$M_{G,\mathrm{NN}}$ & NN predicted Gaia $G$-band absolute magnitude \\
$M_{G,\mathrm{geo}}$ & Observed Gaia $G$-band absolute magnitude derived by {\it Gaia} parallax\\
PARALLAX & Gaia DR3 parallax (mas) before zero point correction\\
PARALLAX\_ERR & Uncertainty in parallax (mas) \\
PARALLAX\_ZPC & Zero point correction of parallax according to \cite{zero_point_correction} \\
EBV & Reddening estimated in this work \\
A\_G & Dust correction value of $M_{G,\mathrm{geo}}$ \\
RUWE & Gaia DR3 RUWE \\
APS & Gaia DR3 ASTROMETRIC\_PARAMS\_SOLVED \\
NEUIA & Gaia DR3 NU\_EFF\_USED\_IN\_ASTROMETRY \\
P\_color & Gaia DR3 PSEUDOcolor\\
ECL\_LAT & Gaia DR3 ECL\_LAT \\
BINARY\_FLAG & Flag of binaries: 1 for single stars, 0 for binaries \\
BINARY\_POSSI\-BILITY & Binary possibility of a star \\
\hline
\end{tabular}
\caption{Contents and columns of our \textsc{SpecDis} VAC product for more than 4 million stars from DESI Year-1 data.}
\label{tab:table}
\end{table*}

\section*{Acknowledgements}
This work is supported by NSFC (12273021,12022307) and the National Key R\&D Program of China (2023YFA1605600, 2023YFA1605601). We thank the sponsorship from Yangyang Development Fund. The computations of this work are carried on the National Energy Research Scientific Computing Center (NERSC) and the Gravity supercomputer at the Department of Astronomy, Shanghai Jiao Tong University. SK acknowledges support from the Science \& Technology Facilities Council (STFC) grant ST/Y001001/1. MV and YW acknowledge support from NASA ATP award (80NSSC20K0509). APC acknowledges support from the Taiwan Ministry of Education Yushan Fellowship, MOE-113-YSFMS-0002-001-P2, and Taiwanese National Science and Technology Council grant 112-2112-M-007-009. HY is supported by T.D. Lee scholarship.

This study was initiated by YW and MV by training the noise free \textsc{sp} model spectra in the Fuji internal data release of DESI, and was passed to SL and WW around the end of 2022 to continue. 

We are grateful for the efforts spent on checking and coordinating our paper by the DESI publication handler, Alejandro Avil\'es. WW is grateful for discussions with Xiaoting Fu, Maosheng Xiang, Xiangxiang Xue, Hao Tian, Jie Wang, Wenda Li, Jundan Nie, Hongliang Yan, Tian Qiu and Yanrui Zhou. 

This paper is based upon work supported by the U.S. Department of Energy (DOE), Office of Science, Office of High-Energy Physics, under Contract No. DE–AC02–05CH11231, and by the National Energy Research Scientific Computing Center, a DOE Office of Science User Facility under the same contract. Additional support for DESI was provided by the U.S. National Science Foundation (NSF), Division of Astronomical Sciences under Contract No. AST-0950945 to the NSF’s National Optical-Infrared Astronomy Research Laboratory; the Science and Technology Facilities Council of the United Kingdom; the Gordon and Betty Moore Foundation; the Heising-Simons Foundation; the French Alternative Energies and Atomic Energy Commission (CEA); the National Council of Humanities, Science and Technology of Mexico (CONAHCYT); the Ministry of Science, Innovation and Universities of Spain (MICIU/AEI/10.13039/501100011033), and by the DESI Member Institutions: \url{https://www.desi.lbl.gov/collaborating-institutions}. Any opinions, findings, and conclusions or recommendations expressed in this material are those of the author(s) and do not necessarily reflect the views of the U. S. National Science Foundation, the U. S. Department of Energy, or any of the listed funding agencies.

The authors are honored to be permitted to conduct scientific
research on Iolkam Du’ag (Kitt Peak), a mountain with particular
significance to the Tohono O’odham Nation.

For the purpose of open access, the author has applied a Creative
Commons Attribution (CC BY) licence to any Author Accepted
Manuscript version arising from this submission.

\appendix

\section{Examples of Principal Components}

\begin{figure}
\begin{center}
\includegraphics[width=0.49\textwidth]{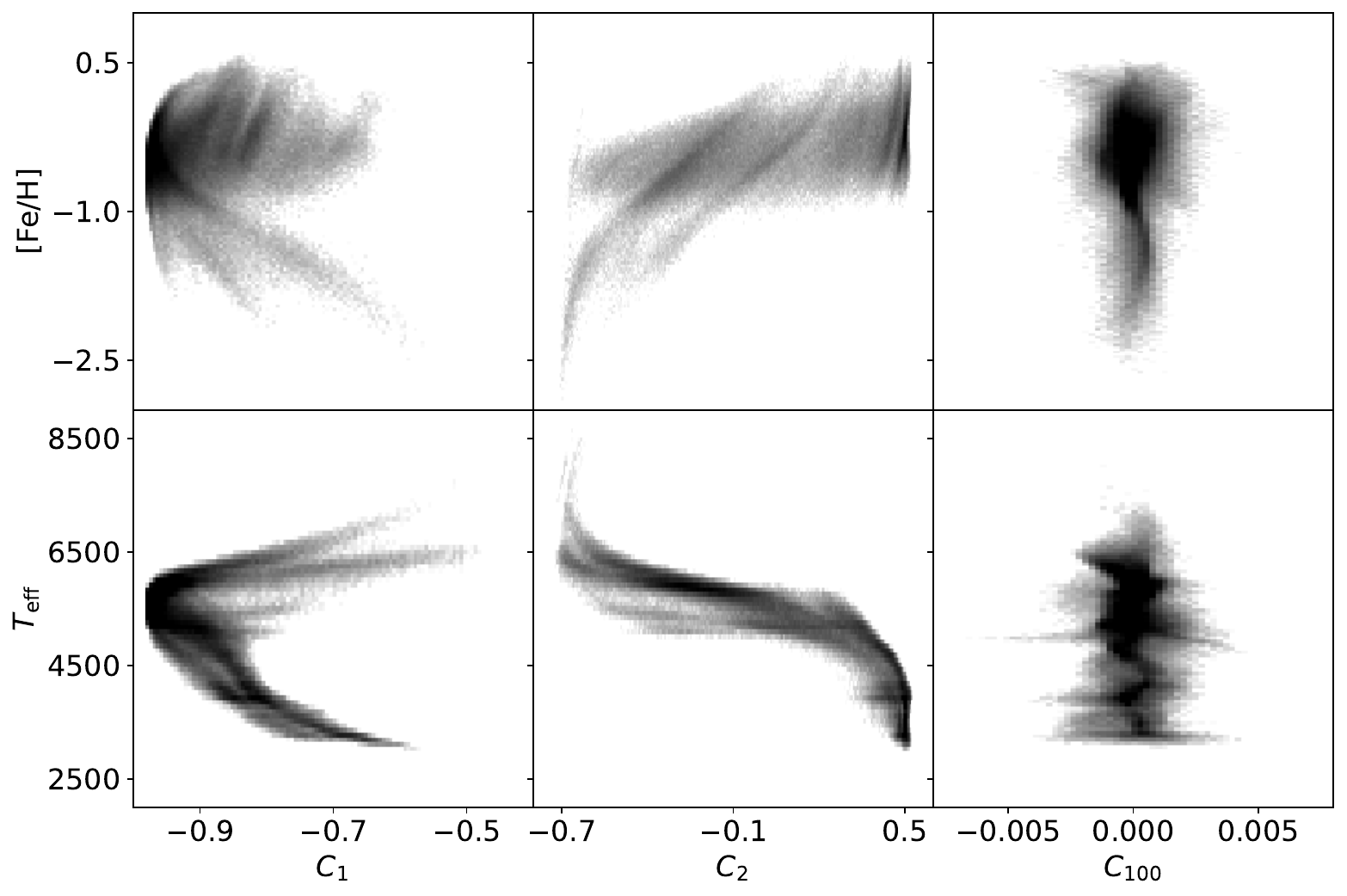}%
\end{center}
\caption{The relationships between [Fe/H], $T_\mathrm{eff}$, and the coefficients of three principal components. $C_1$, $C_2$ and $C_{100}$ represent the coefficients of the first, second and 100-th PCs. Here [Fe/H] and $T_\mathrm{eff}$ are from \textsc{rvs} table.}
\label{fig:pc amplitude}
\end{figure}

\begin{figure*} 
\begin{center}
\includegraphics[width=0.8\textwidth,height=0.8\textheight]{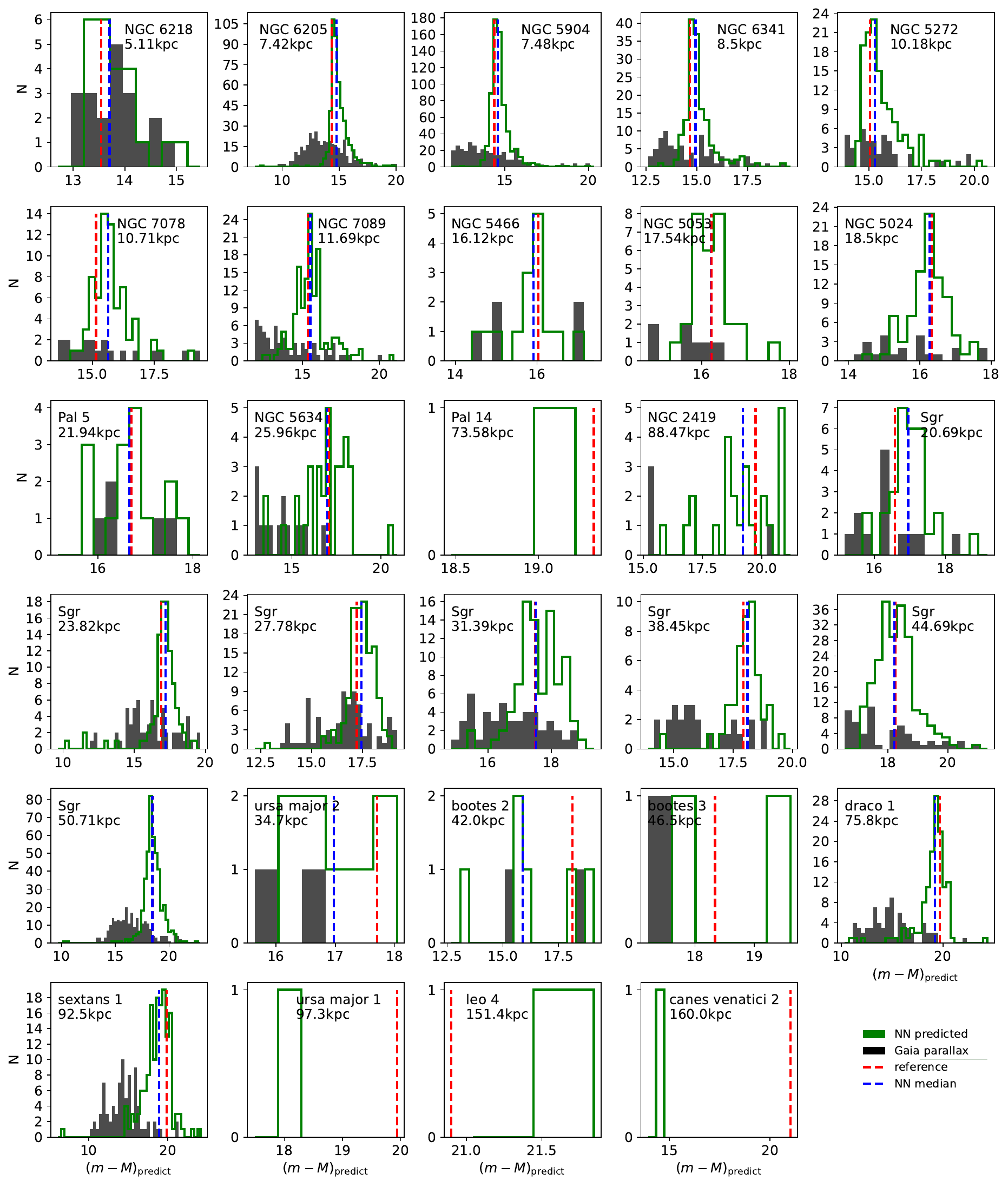}%
\end{center}
\caption{The distributions of $(m-M)_\mathrm{predict}$ predicted by NN (the green histogram) and $(m-M)_\mathrm{geo}$ derived from {\it Gaia} parallax (the black filled histogram). Each panel shows an individual GC, dwarf galaxy or a given distance bin of the Sagittarius stream. The red vertical dashed line marks the reference distance modulus of the parent object and the blue vertical dashed line marks the median of distance modulus predicted by NN. When the number of member stars is too few, we do not show the blue vertical line. The text in each panel indicates the name of the object and its distance.}
\label{fig:overestimated gc}
\end{figure*}

\begin{table}
\centering
\caption{Distances for GCs and dwarf galaxies, the member stars of which are used for distance external validations in this paper. Here those GCs and dwarf galaxies having fewer than 10 member stars not shown. $d_\mathrm{ref}$ and $d_\mathrm{NN}$ are the reference distance and the median value of the NN predicted distances. a and b stand for reference source of \citep{2021MNRAS.505.5957B} and \citep{2022ApJ...940..136P}.}
\begin{tabular}{lccccccc}
\hline
\textbf{Name} &  \textbf{$d_\mathrm{ref}$ [kpc]} & \textbf{$d_{\mathrm{NN}}$ [kpc]} & \textbf{References} & \textbf{Name} &  \textbf{$d_\mathrm{ref}$ [kpc]} & \textbf{$d_{\mathrm{NN}}$ [kpc]} & \textbf{References} \\
\hline
NGC 6218 &  $5.109^{+0.049}_{-0.048}$   & 5.50 & a & NGC 5053 &  $ 17.537^{+0.235}_{-0.232}$  & 17.5 & a\\
NGC 6205 &  $ 7.419^{+0.076}_{-0.075}$  & 8.99 & a & NGC 5024 &  $ 18.498^{+0.185}_{-0.183}$  & 17.9 & a\\
NGC 5904 &  $ 7.479^{+0.060}_{-0.060}$  & 8.28 & a & Pal5 &  $ 21.941^{+0.520}_{-0.508}$  & 21.5 & a\\
NGC 6341 &  $ 8.501^{+0.071}_{-0.070}$  & 9.69 & a & NGC 5634 &  $ 25.959^{+0.628}_{-0.613}$  & 25.0 & a\\
NGC 5272 &  $ 10.175^{+0.082}_{-0.081}$ & 11.4 & a & NGC 2419 &  $  88.471^{+2.437}_{-2.371}$  & 69.2 & a\\
NGC 7078 &  $ 10.709^{+0.096}_{-0.095}$ & 13.4 & a & draco 1 & $  75.8^{+5.4}_{-5.4}$ & 70.1 & b\\
NGC 7089 &  $ 11.693^{+0.115}_{-0.114}$ & 12.6 & a & sextans 1 & $  92.5^{+2.5}_{-2.5}$ & 59 & b\\
NGC 5466 &  $ 16.120^{+0.164}_{-0.162}$ & 15.2 & a\\
\hline
\end{tabular}
\label{tab:validation_table}
\end{table}

Figure~\ref{fig:pc amplitude} illustrates the correlation between the amplitudes of the first two PCs ($C_1$ and $C_2$) and the 100-th PC ($C_\mathrm{100}$) versus the iron abundance ([Fe/H]) and effective temperatures ($T_\mathrm{eff}$) of the stars. The distribution of $C_1$ and $C_2$ indicates that the coefficients of PCs are strongly correlated with these stellar properties, suggesting that they contain physical information. The two right panels based on $C_\mathrm{100}$ show that for higher order PCs the correlations with stellar properties are significantly decreased. 

However, $C_1$ and $C_2$ do not appear to correlate with [Fe/H] and $T_\mathrm{eff}$ linearly, and the trend is not monotonic. In the two left panels, with the increase in [Fe/H] or $T_\mathrm{eff}$, $C_1$ first decreases and then increases. In addition, there are clearly sub features. This likely reflects that $C_1$ and $C_2$ do not contain the information of only one stellar parameter, but can be correlated with the combinations of a few different parameters. This has been explored in, for example, a previous study of \cite{2012MNRAS.421..314C}. Understanding the physical meanings of the amplitudes for different PCs, however, is beyond the scope of this current paper.

\section{External validation of member stars in GCs, dwarf galaxies and the Sgr}

Figure~\ref{fig:overestimated gc} shows the external validations of our distances, based on the member stars of each individual GC, dwarf galaxy and the Sgr stream (see Section~\ref{sec:validation set}). The green and black filled histograms show the distributions our predicted distance modulus ($(m-M)_\mathrm{predict}$) and the distance modulus calculated from {\it Gaia} parallax ($(m-M)_\mathrm{geo}$), respectively. 
The red vertical dashed line marks the reference distance of the parent GC, dwarf galaxy or stream \citep{2021MNRAS.505.5957B,2022ApJ...940..136P}. 
We divide Sgr member stars to six bins of different distance ranges, and the red vertical dashed line is the mean distance modulus of each bin. The blue vertical dashed line in each panel marks the median distance modulus based on our measurements. Our measured distances are approximately unbiased and show significantly smaller scatters, compared with the distances from {\it Gaia} parallax beyond 7~kpc, when there are enough numbers of member stars. The slight overestimated distances of NGC 6205 and NGC 5904 observed here contributes to the bias noticed in Figure~\ref{fig:validation}. For distant systems, our measurements show some under estimates, consistent with Figure~\ref{fig:validation}. Note when there is a large discrepancy between the blue and red vertical dashed lines, it is mainly because the number of matched member stars is too few. We provide the reference distances and the median distance of our NN prediction, for different GCs and dwarf galaxies used for the external validation, in Table~\ref{tab:validation_table}.

\vspace{1em}

\bibliography{master}{}
\bibliographystyle{aasjournal}
\end{document}